\documentclass[12pt,aps,tightenlines]{revtex4}
\pdfoutput=1

\usepackage{amsmath}
\usepackage{amsfonts}
\usepackage{bm}
\usepackage{graphicx}
\usepackage{subfigure}
\usepackage{color}

\graphicspath{{figs_pdf/}}

\newcommand{\Small}{\delta}
\newcommand{\Stresstensor}{T}

\newcommand{\Cn}{C_{\mathrm{n}}}

\newtheorem{prop}{Proposition}

\begin{document}
%+Title
\title{Nonlinear dynamics of phase separation in thin films}
\author{Lennon \'O N\'araigh}
\affiliation{Department of Chemical Engineering, Imperial College
  London, SW7 2AZ, United Kingdom}
\author{Jean-Luc Thiffeault}
\email{jeanluc@mailaps.org}
\affiliation{Department of Mathematics, University of Wisconsin, Madison,
WI 53706, USA}
\date\today
%-Title
%
\begin{abstract}
  We present a long-wavelength approximation to the Navier--Stokes
  Cahn--Hilliard equations to describe phase separation in thin films.
  The equations we derive underscore the coupled behaviour of
   free-surface variations and phase separation.  
  We introduce a repulsive substrate-film interaction potential and analyse   the
  resulting fourth-order equations by constructing a Lyapunov functional,   which, combined with the regularizing repulsive potential, gives rise to   a positive lower bound for the free-surface height.
  The value of this lower bound depends on the parameters of the problem,   a result which we compare with numerical simulations.  While the theoretical   lower bound is an obstacle to the rupture of a film that initially is    everywhere of finite height, it is 
  not sufficiently sharp to represent accurately the parametric
  dependence of the observed dips or `valleys' in free-surface height.  We   observe these valleys across zones where the concentration of the binary   mixture   changes   sharply, indicating the formation of bubbles.  Finally,   we carry out numerical simulations without the repulsive interaction, and find that the film ruptures in finite time, while the gradient of the Cahn--Hilliard concentration develops a singularity.
\end{abstract}

\maketitle
\section{Introduction}
\label{sec:intro} 

\noindent Below a certain critical temperature, a well-mixed binary
fluid spontaneously separates into its component parts, forming
domains of pure liquid.  This process can be characterised by the
Cahn--Hilliard equation, and numerous studies describe the physics and
mathematics of phase separation~\cite{CH_orig,Bray_advphys,Elder1988,Zhu_numerics,ONaraigh2006}.  In this paper we study phase
separation in a thin layer, in which the varying free-surface and
concentration fields are coupled through a pair of nonlinear evolution
equations.

Cahn and Hilliard introduced their eponymous equation
in~\cite{CH_orig} to model phase separation in a binary alloy.  Since
then, the model has been used in diverse applications: to describe
polymeric fluids~\cite{Aarts2005}, fluids with interfacial
tension~\cite{Spelt2007,LowenTrus}, and self-segregating populations in
biology~\cite{Murray1981}.
An analysis of the Cahn--Hilliard (CH) equation was given by Elliott and
Zheng~\cite{Elliott_Zheng}, where they obtain existence,
uniqueness, and regularity results.  Several authors have
developed generalisations of the CH equation: a variable-mobility
model was introduced by Elliott and Garcke~\cite{Elliott_varmob},
while nonlocal effects were considered by Gajweski and
Zacharias~\cite{Gajewski_nonlocal}.  These additional features do not
qualitatively change the phase separation, and we therefore turn to
one mechanism that does: the coupling of a flow field to the
Cahn--Hilliard equation~\cite{Bray_advphys}.
In this case, the Cahn--Hilliard concentration equation is modified by an
advection term, and the flow field is either prescribed or evolves according
to some fluid equation.  Ding and co-workers~\cite{Spelt2007} provide
a derivation of coupled Navier--Stokes Cahn--Hilliard (NSCH) equations in
which the velocity advects the phase-separating concentration field, while
concentration gradients modify the velocity through an additional stress
term in the momentum equation.  A similar model has been produced by Lowengrub and Truskinowsky~\cite{LowenTrus}.  Such models have formed the basis of numerical
studies of binary fluids~\cite{Berti2005}, while other studies without
this feedback term highlight different regimes of phase separation under
flow~\cite{chaos_Berthier, ONaraigh2006, Lacasta1995}.  Here, the
NSCH equations form the starting point for our asymptotic analysis.

As in other applications involving the Navier--Stokes equations, the
complexity of the problem is reduced when the fluid is spread thinly
on a substrate, and the upper vertical boundary forms a free
surface~\cite{Oron1997,Matar2009}.  Then, provided lateral gradients are small
compared to vertical gradients, a long-wavelength approximation is
possible, in which the full equations with a moving boundary at the
free surface are reduced to a single equation for the free-surface
height.  In the present case, the reduction yields two equations: one
for the free surface, and one for the Cahn--Hilliard concentration.
The resulting thin-film Stokes Cahn--Hilliard equations have already
been introduced by the authors in~\cite{ONaraigh2007}, although the
focus there was on control of phase separation and numerical
simulations in three dimensions.  Here we confine ourselves to the
two-dimensional case: we derive the thin-film equations from first
principles, present analysis of the resulting equations, and highlight
the impossibility of film rupture, once a regularizing potential is prescribed.  We use numerical simulations to show that in the absence of this regularizing potential, the film does indeed rupture, an event that coincides with the development of a singularity in the concentration gradient.
 
Along with the simplification of the problem that thin-film theory
provides, there are many practical reasons for studying phase
separation in thin layers.  Thin polymer films are used in the
fabrication of semiconductor devices, for which detailed knowledge of
film morphology is required~\cite{Karim2002}.  Other industrial
applications of polymer films include paints and coatings, which are
typically mixtures of polymers.  One potential application of the
thin-film Cahn--Hilliard theory is in self-assembly~\cite{Krausch1994,
  Xia2004, Lu2005, Putkaradze2005, Kim2006}.  Here molecules (usually
residing in a thin layer) respond to an energy-minimisation
requirement by spontaneously forming large-scale structures.
Equations of Cahn--Hilliard type have been proposed to explain the
qualitative features of self-assembly~\cite{Putkaradze2005,Holm2005},
and knowledge of variations in the film height could enhance these
models.  Indeed in~\cite{ONaraigh2007} the authors use the present
thin-film Cahn--Hilliard model in three dimensions to control phase
separation, a useful tool in applications where it is necessary for
the molecules in the film to form a given structure.

The mathematical analysis of thin-film equations was given great impetus by Bernis
and Friedman in~\cite{Friedman1990}.  They focus 
on the basic thin-film equation,
\begin{equation}
\frac{\partial h}{\partial t}=-\frac{\partial}{\partial x}\left(h^n\frac{\partial^3{h}}{\partial{x^3}}\right),
\label{eq:basic_h}
\end{equation}
with no-flux boundary conditions on a line segment, and smooth
nonnegative initial conditions.  For $n=1$ this equation describes a
thin bridge between two masses of fluid in a Hele--Shaw cell, for
$n<3$ it is used in slip models as $h\rightarrow0$~\cite{Myers1998},
while for $n=3$ it gives the evolution of the free surface of a thin
film experiencing capillary forces~\cite{Oron1997}.
Using a decaying free-energy functional, they analyzed Eq.~\eqref{eq:basic_h} and obtained results concerning the existence, uniqueness, and regularity of solutions, as a function of the exponent $n$.  Only for $n\geq4$ does a classical, smooth solution exist; this fact is established by construction of an entropy functional.
 This paper~\cite{Friedman1990} has inspired
other work on the subject~\cite{Bertozzi1996,
  Bertozzi1998,Laugesen2002}, in which the effect of a Van der Waals
term on Eq.~\eqref{eq:basic_h} is investigated.  These works provide
results concerning regularity, long-time behaviour, and film rupture
in the presence of an attractive Van der Waals force.
More relevant to the present work is the paper by Wieland and
Garcke~\cite{Wieland2006}, in which a pair of partial differential
equations describes the coupled evolution of free-surface variations
and surfactant concentration.  The authors derive the relevant
equations using the long-wavelength theory, obtain a decaying energy
functional, and prove results concerning the existence and
non-negativity of solutions.

When the binary fluid forms a thin film on a substrate, we shall show
in Sec.~\ref{sec:model} that a long-wave approximation simplifies the density- and viscosity-matched
Navier--Stokes Cahn--Hilliard equations, which reduce to a pair of
coupled evolution equations for the free surface and concentration.
If $h(x,t)$ is the scaled free-surface height, and
$c(x,t)$ is the binary fluid concentration, then the
dimensionless equations take the form
\begin{subequations}
\begin{equation}
\frac{\partial h}{\partial t}+\frac{\partial J}{\partial x}=0,\qquad
\frac{\partial}{\partial t}\left(h c\right)+\frac{\partial}{\partial x}\left(Jc\right)=\frac{\partial}{\partial{x}}\left(h\frac{\partial\mu}{\partial{x}}\right),
\end{equation}
where
\begin{equation}
J = \tfrac{1}{2} h^2 \frac{\partial\sigma}{\partial{x}}
- \tfrac{1}{3} h^3 \bigg\{
           \frac{\partial}{\partial{x}}
           \left(-\frac{1}{C}\frac{\partial^2{h}}{\partial{x}^2}
             + \phi\right) + \frac{r}{h} \frac{\partial}{\partial{x}}
           \left[h{\left(\frac{\partial{c}}{\partial{x}}
             \right)}^2\right]\bigg\},
\end{equation}
\begin{equation}
\mu=c^3-c-\frac{\Cn^2}{h}\frac{\partial}{\partial{x}}\left(h\frac{\partial{c}}{\partial{x}}\right).
\end{equation}%
\label{eq:model_intro}%
\end{subequations}%
Here $C$ is the capillary number, $r$ measures the strength of
coupling between the concentration and free-surface variations
(backreaction), and $\Cn$ is the scaled interfacial
thickness --- sometimes called the Cahn number.  Additionally, $\sigma$ is the dimensionless,
spatially-varying surface tension, and $\phi$ is the body-force
potential acting on the film.  In this paper we focus on a class of potentials that models a repulsive interaction between the film and the substrate, thus preventing rupture.   This enables us to focus on late-time phase separation, which is synonymous with a tendency to equilibrium.
However, rupture is in itself an important
feature in thin-film equations~\cite{Oron1997, Bertozzi1996, Bertozzi1998}: we therefore use numerical methods to highlight the possibility of rupture in the absence of a repulsive film-substrate interaction.

The paper is organised as follows.  In Sec.~\ref{sec:model} we discuss
the Navier--Stokes Cahn--Hilliard equation and the scaling laws that
facilitate the passage to the long-wavelength equations, and we derive
Eq.~\eqref{eq:model_intro}.  In Sec.~\ref{sec:analysis} we perform linear and non-linear analyses of the long-wavelength equations.  Our non-linear analysis centres on finding a Lyapunov functional for a given class of potentials.  We derive \textit{a priori} bounds for a given positive ($h>0$) solution, and estimate the minimum value of the free-surface height.
  In Sec.~\ref{sec:height_dip} we outline a series of numerical studies, with and without a regularizing potential, and we discuss the
dependence of the minimum free-surface height on the problem
parameters.  Finally, in Sec.~\ref{sec:conclusions} we present our
conclusions.

\section{The model equations: derivation}
\label{sec:model}
\noindent 
In this section we introduce the two-dimensional Navier--Stokes
Cahn--Hilliard (NSCH) equation set.  We focus on the so-called matched case, wherein both components of the binary mixture have the same density and viscosity.
We discuss the assumptions underlying the long-wavelength
approximation.
We enumerate the scaling rules necessary to obtain the simplified
equations.  
Finally, we arrive at a set of equations that describe phase separation in
a thin film subject to arbitrary body forces.

The full NSCH equations describe the coupled effects of phase separation
and flow in a binary fluid.  If the fluids are density- and viscosity matched, then the models described in the references~\cite{Spelt2007,LowenTrus} agree; this is the case we study.  If $\bm{v}$ is the fluid velocity and $c$ is
the concentration of the mixture, where $c=\pm1$ indicates total segregation,
then these fields evolve as
\begin{subequations}
\begin{gather}
        \frac{\partial \bm{v}}{\partial t}+\bm{v}\cdot\nabla\bm{v}=\nabla\cdot\Stresstensor-\frac{1}{\rho}\nabla\phi,\\
        \frac{\partial c}{\partial t}+\bm{v}\cdot\nabla c=D \nabla^2\left(c^3-c-\gamma\nabla^2 c\right),\\
        \nabla\cdot\bm{v}=0,
\end{gather}%
\label{eq:NSCH}%
\end{subequations}%
where
\begin{equation}
        \Stresstensor_{ij} =-\frac{p}{\rho}\delta_{ij}+\nu\left(\frac{\partial
        v_i}{\partial
        x_j}+\frac{\partial
        v_j}{\partial x_i}\right)-\beta\gamma\frac{\partial c}{\partial x_i}\frac{\partial
        c}{\partial x_j}
\label{eq:NSCH_tensor}%
\end{equation}%
is the stress tensor, $p$ is the fluid pressure, $\phi$ is the body
potential and $\rho$ is the constant density.  The constant $\nu$ is
the kinematic viscosity, $\nu=\eta/\rho$, where $\eta$ is the dynamic
viscosity.  Additionally, $\beta$ is a constant with units of
$[\mathrm{Energy}][\mathrm{Mass}]^{-1}$, $\sqrt{\gamma}$ is a constant
that gives the typical width of interdomain transitions, and $D$ a
diffusion coefficient with dimensions
$[\mathrm{Length}]^2[\mathrm{Time}]^{-1}$.

We impose the following boundary conditions (BCs).  On the lateral boundaries ($x$-direction), the velocity must satisfy the no-slip condition, while $c_x$ and $c_{xxx}$ must vanish too.  Alternatively, we may enforce periodicity, and demand that the velocity, $c$, $c_x$, and $c_{xx}$ be periodic functions of the lateral coordinate.
Finally, if the system has a free surface in the vertical or $z$-direction, then the vertical BCs are the following:
\begin{subequations}
\begin{equation}
u = w = c_z = c_{zzz}\text{ on }z=0,
\end{equation}
while on the free surface $z=h(x,t)$ they are
\begin{equation}
\hat{n}_i \hat{n}_j \Stresstensor_{ij} = -\sigma\kappa,\qquad \hat{n}_i \hat{t}_j
\Stresstensor_{ij} = -\frac{\partial\sigma}{\partial
s},
\label{eq:BC_stress}
\end{equation}
\begin{equation}
w=\frac{\partial h}{\partial t}+u\frac{\partial h}{\partial x},
\label{eq:BC_height}
\end{equation}
\begin{equation}
\hat{n}_i \partial_i c = 0, \qquad \hat{n}_i \partial_i \nabla^2 c = 0,
\label{eq:BC_c}
\end{equation}%
\label{eq:BCs}%
\end{subequations}%
where $\hat{\bm{n}} =
(-\partial_x{h}\,,\,1)/{[{1+(\partial_x{h})^2}]^{1/2}}$ is the unit
normal to the surface, $\hat{\bm{t}}$ is the unit vector tangent to
the surface, $s$ is the surface coordinate, $\sigma$ is the surface
tension, and $\kappa$ is the mean curvature,
\begin{equation*}
\kappa=\nabla\cdot\hat{\bm{n}}
=\frac{\partial_{xx}h}{\left[1+\left(\partial_x{h}\right)^2\right]^{\frac{3}{2}}};
\end{equation*}
these free-surface conditions are standard and are discussed in the review papers~\cite{Oron1997,Matar2009}.
This choice of BCs guarantees the conservation of the total mass and volume,
\begin{equation}
\text{Mass} = \int_{\text{Dom}(t)} dxdz\,c(x,z,t),\qquad
\text{Volume} = \int_{\text{Dom}(t)}dxdz.
\label{eq:integral_quantities}
\end{equation}
Here 
$\text{Dom}(t)%=\{\left(x,y\right)|x\in\left[0,L\right],z\in\left[0,h(x,t)\right]\}
$
represents the time-dependent domain of integration, owing to the
variability of the free-surface height.  Note that in view of the
concentration BC~\eqref{eq:BC_c}, the stress BC~\eqref{eq:BC_stress}
and does not contain $c(\bm{x},t)$ or its derivatives.

These equations simplify considerably if the fluid forms a thin layer
of mean thickness $h_0$, for then the scale of lateral variations
$\ell$ is large compared with the scale of vertical variations
$h_0$.  Specifically, the parameter $\Small = h_0/\ell$ is small,
and after nondimensionalisation of Eq.~\eqref{eq:NSCH} we expand its
solution in terms of this parameter, keeping only the lowest-order
terms.  For a review of this method and its applications,
see~\cite{Oron1997,Matar2009}.  For simplicity, we shall work in two dimensions,
but the generalisation to three dimensions is easily
effected~\cite{ONaraigh2007}.

In terms of the small parameter $\Small$, the equations
nondimensionalise as follows.  The diffusion time scale is $t_0 =
\ell^2 / D = h_0^2 / \left(\Small^2 D\right)$ and we choose this to
be the unit of time.  Then the unit of horizontal velocity is $u_0 =
\ell/ t_0 = \Small D / h_0$ so that $u = \left(\Small D /
  h_0\right)U$, where variables in upper case denote dimensionless
quantities.  Similarly, the vertical velocity is $w = \left(\Small^2 D / h_0\right)W$, and the free-surface height is $h=h_0 H$.  The dimensionless coordinates are introduced through the equations $x=\ell X$, $z=h_0Z$.  Finally, for the equations of motion to be half-Poiseuille
at $O\left(1\right)$ (in the absence of the backreaction) we choose $p
= \left(\eta D / h_0^2\right)P$ and $\phi = \left(\eta D /
  h_0^2\right)\Phi$.  
We also stipulate the following form for the surface tension:
\[
\sigma=\sigma_0\left[1+\Small^q f\left(X\right)\right],
\]  
where the exponent $q$ is yet to be determined.
Using these scaling rules, the dimensionless
momentum equations are
\begin{multline}
\Small Re \left(\frac{\partial U}{\partial T} + U\frac{\partial
U}{\partial X}+ W\frac{\partial U}{\partial Z}\right) 
= -\frac{\partial}{\partial X}\left(P+\Phi\right)+\Small^2\frac{\partial^2
U}{\partial X^2}+\frac{\partial^2 U}{\partial Z^2} \\
-\tfrac{1}{2}\frac{\beta\gamma}{\nu D}\frac{\partial}{\partial
X}\bigg[\Small^2\left(\frac{\partial c}{\partial X}\right)^2
+\left(\frac{\partial c}{\partial Z}\right)^2\bigg]-\frac{\beta\gamma}{\nu
D}\frac{\partial c}{\partial X}\bigg[\Small^2\frac{\partial^2 c}{\partial X^2}+\frac{\partial^2 c}{\partial Z^2}\bigg],
\label{eq:thin_film1}
\end{multline}
\begin{multline}
\Small^3 Re \left(\frac{\partial W}{\partial T} + U\frac{\partial
W}{\partial X} + W\frac{\partial W}{\partial Z}\right) 
= -\frac{\partial}{\partial Z}\left(P+\Phi\right)+\Small^4\frac{\partial^2
W}{\partial X^2}+\Small^2\frac{\partial^2 W}{\partial Z^2}  \\
-\tfrac{1}{2}\frac{\beta\gamma}{\nu D}\frac{\partial}{\partial
Z}\bigg[\Small^2\left(\frac{\partial c}{\partial X}\right)^2+\left(\frac{\partial
c}{\partial Z}\right)^2\bigg]
-\frac{\beta\gamma}{\nu D}\frac{\partial c}{\partial Z}\bigg[\Small^2\frac{\partial^2
c}{\partial X^2}+\frac{\partial^2 c}{\partial Z^2}\bigg],
\label{eq:thin_film2}
\end{multline}
\begin{equation}
\frac{\partial U}{\partial X}+\frac{\partial W}{\partial Z}=0,
\label{eq:cont}
\end{equation}
where
\begin{equation}
Re = \frac{u_0h_0}{\nu} = \frac{\Small D }{\nu} =
O\left(1\right).
\end{equation}
The choice of ordering for the Reynolds number $Re$ allows us to recover
half-Poiseuille flow at $O\left(1\right)$.  We delay choosing the ordering
of the dimensionless group $\beta\gamma/D\nu$ until we have examined the
concentration equation, which in nondimensional form is
\begin{multline}
\Small^2 \left(\frac{\partial c}{\partial T} + U\frac{\partial c}{\partial
X} + W\frac{\partial c}{\partial Z}\right)\\
= \Small^2 \frac{\partial^2}{\partial X^2}\left(c^3-c\right)+\frac{\partial^2}{\partial Z^2}\left(c^3-c\right)
-\Small^4 \Cn^2\frac{\partial^4c}{\partial X^4}-\Cn^2\frac{\partial^4
c}{\partial Z^4}
-2\Small^2 \Cn^2 \frac{\partial^2}{\partial X^2}\frac{\partial c}{\partial  Z^2},
\label{eq:conc0}
\end{multline}
where $\Cn^2=\gamma/h_0^2$.  By switching off the
backreaction in the momentum equations (corresponding to
$\beta\gamma/D\nu\rightarrow0$), we find the trivial solution to
the momentum equations, $U = W = \partial_X\left(P+\Phi\right)
= \partial_Z\left(P+\Phi\right) = 0$, $H = 1$.  The concentration
boundary conditions are then $c_Z = c_{ZZZ} = 0$ on $Z = 0,1$, which
forces $c_Z\equiv0$ so that the Cahn--Hilliard equation is simply
\[
\frac{\partial c}{\partial T} = \frac{\partial^2}{\partial X^2}\left(c^3-c\right)-\Small^2 \Cn^2\frac{\partial^4c}{\partial
X^4}.
\]
To make the lubrication approximation consistent, we take 
\begin{equation}
\Small\Cn= \widetilde{\Cn}= \Small\sqrt{\gamma} /
h_0 = O\left(1\right).
\label{eq:gamma_h}
\end{equation}
We now carry out a long-wavelength approximation to
Eq.~\eqref{eq:conc0}, writing $U=U_0+O\left(\Small\right)$,
$W=W_0+O\left(\Small\right)$, $c=c_0+\Small c_1+\Small^2 c_2+\cdots$.
We examine the boundary conditions on $c(\bm{x},t)$ first.
They are $\hat{\bm{n}}\cdot\nabla c =
\hat{\bm{n}}\cdot\nabla\nabla^2c=0$ on $Z=0,H$; on $Z=0$ these
conditions are simply $\partial_Z c = \partial_{ZZZ} c=0$, while on
$Z=H$ the surface derivatives are determined by the relations
\[
\hat{\bm{n}}\cdot\nabla\ \propto\ -\Small^2 H_X\partial_X+\partial_Z,
\]
\[
\hat{\bm{n}}\cdot\nabla\nabla^2\ \propto\ -\Small^4
H_X\partial_{XXX}-\Small^2H_X\partial_X\partial_{ZZ}+\Small^2\partial_{XX}\partial_Z+\partial_{ZZZ}.
\]
Thus, the BCs on $c_0$ are simply $\partial_Z c_0 = \partial_{ZZZ} c_0=0$
on $Z=0,H$, which forces $c_0=c_0\left(X,T\right)$.  Similarly, we
find  $c_1=c_1\left(X,T\right)$, and
\[
\frac{\partial c_2}{\partial Z}=Z\frac{H_X}{H}\frac{\partial
c_0}{\partial X},\qquad
\frac{\partial^2 c_2}{\partial Z^2}=\frac{H_X}{H}\frac{\partial
c_0}{\partial X}, \qquad \text{for any }Z\in \left[0,H\right].
\]
In the same manner, we derive the results $\partial_{ZZZZ}c_2=\partial_{ZZZZ}c_3=0$.
Using these facts, Eq.~\eqref{eq:conc0} becomes
\begin{multline*}
\frac{\partial c_0}{\partial T} + U_0\frac{\partial c_0}{\partial
X} =
\\
\frac{\partial^2}{\partial X^2}\left(c_0^3-c_0\right)
-\widetilde{\Cn}^2\frac{\partial^4c}{\partial X^4}
+\left(3c_0^2-1\right)\frac{H_X}{H}\frac{\partial c_0}{\partial X}
-2\widetilde{\Cn}^2\frac{\partial^2}{\partial X^2}\frac{H_X}{H}\frac{\partial
c_0}{\partial X}
-\widetilde{\Cn}\frac{\partial^4 c_4}{\partial Z^4}.
\end{multline*}
We now integrate this equation from $Z=0$ to $H$ and use the boundary conditions
\[
\frac{\partial^3 c_4}{\partial Z^3}=0\quad\text{on}\quad Z=0,
\]
\begin{equation*}
\frac{\partial^3 c_4}{\partial Z^3}=
H_X\frac{\partial^3 c_0}{\partial X^3}+
H_X\frac{\partial}{\partial X}\left(\frac{H_X}{H}\frac{\partial c_0}{\partial
X}\right)
-H\frac{\partial^2}{\partial X^2}\left(\frac{H_X}{H}\frac{\partial c_0}{\partial
X}\right)\quad\text{on}\quad Z=H.
\end{equation*}
After rearrangement, the concentration equation becomes
\begin{multline*}
H\frac{\partial c_0}{\partial T}+H\langle U_0\rangle\frac{\partial c_0}{\partial
X} = 
\\
H\frac{\partial^2}{\partial X^2}\left[c_0^3-c_0-\widetilde{\Cn}^2\frac{\partial^2
c_0}{\partial X^2}-\widetilde{\Cn}^2\frac{H_X}{H}\frac{\partial c_0}{\partial
X}\right]
+\frac{\partial H}{\partial X}\frac{\partial}{\partial X}\left[c_0^3-c_0-\widetilde{\Cn}^2\frac{\partial^2
c_0}{\partial X^2}-\widetilde{\Cn}^2\frac{H_X}{H}\frac{\partial c_0}{\partial
X}\right],
\end{multline*}
where
\[
\langle U_0\rangle = \frac{1}{H}\int_0^H U_0\left(X,Z,T\right)dZ
\]
is the vertically-averaged velocity.  Introducing
\[
\mu = c_0^3-c_0-\frac{\widetilde{\Cn}^2}{H}\frac{\partial}{\partial X}\left(H\frac{\partial c_0}{\partial
X}\right),
\]
the thin-film Cahn--Hilliard equation becomes
\begin{equation}
\frac{\partial c_0}{\partial T}+\langle U_0\rangle\frac{\partial c_0}{\partial
X} = \frac{1}{H}\frac{\partial}{\partial X}\left(H\frac{\partial\mu}{\partial
X}\right).
\label{eq:conc_eqn_ornament}
\end{equation}

We are now able to perform the long-wavelength approximation to Eqs.~\eqref{eq:thin_film1}
and~\eqref{eq:thin_film2}.  At lowest order, Eq.~\eqref{eq:thin_film2} is~$\partial_ Z\left(P+\Phi\right)=0$,
since~$c_0=c_0(X,T)$, and hence
\[
P+\Phi = P_{\mathrm{surf}}+\Phi_{\mathrm{surf}}\equiv P\left(X,H(X,T),T\right)+\Phi\left(X,H(X,T),T\right).
\]
We introduce a dimensionless group to measure the strength of the interaction between the concentration and velocity fields; we also specify its order of magnitude:
\begin{equation}
r=\frac{\Small^2\beta\gamma}{D\nu}=O\left(1\right).
\end{equation}
Later on we refer to this quantity as the `backraction strength', since it is a measure of the extent to which concentration gradients feed back into the flow field.  Using this dimensionless group, Eq.~\eqref{eq:thin_film1} becomes
\[
\frac{\partial^2 U_0}{\partial Z^2}=\frac{\partial}{\partial X}\left(P_{\mathrm{surf}}+\Phi_{\mathrm{surf}}\right)+r\frac{\partial}{\partial
X}\left(\frac{\partial c_0}{\partial X}\right)^2+r\frac{\partial c_0}{\partial
X}\frac{\partial^2 c_2}{\partial Z^2}.
\]
Using $\partial_{ZZ}c_2=\left(H_X/H\right)\left(\partial c_0/\partial
X\right)$ this becomes
\begin{equation}
\frac{\partial^2 U_0}{\partial Z^2}=\frac{\partial}{\partial X}\left(P_{\mathrm{surf}}+\Phi_{\mathrm{surf}}\right)+\frac{r}{H}\frac{\partial}{\partial
X}\left[H\left(\frac{\partial c_0}{\partial X}\right)^2\right].
\label{eq:d2U}
\end{equation}
At lowest order, the BC~\eqref{eq:BC_stress} reduces to
\begin{equation}
\frac{\partial U_0}{\partial
  Z}=\frac{\partial\Sigma}{\partial{X}}\quad\text{on}\quad Z=H,
\label{eq:sigma_x}
\end{equation}
which combined with Eq.~\eqref{eq:d2U} yields the relation
\[
\frac{\partial U_0}{\partial Z}=\frac{\partial\Sigma}{\partial X}+\left(Z-H\right)\bigg\{
\frac{\partial}{\partial X}\left(P_{\mathrm{surf}}+\Phi_{\mathrm{surf}}\right)+\frac{r}{H}\frac{\partial}{\partial
X}\left[H\left(\frac{\partial c_0}{\partial X}\right)^2\right]
\bigg\}.
\]
Here $\Sigma$ is the dimensionless, spatially-varying component of the surface tension.  Making
use of the BC $U_0=0$ on $Z=0$ and integrating again, we obtain the result
\begin{equation}
U_0\left(X,Z,T\right)=Z\frac{\partial\Sigma}{\partial X}+\left(\tfrac{1}{2}Z^2-HZ\right)\bigg\{
\frac{\partial}{\partial X}\left(P_{\mathrm{surf}}+\Phi_{\mathrm{surf}}\right)+\frac{r}{H}\frac{\partial}{\partial
X}\left[H\left(\frac{\partial c_0}{\partial X}\right)^2\right]
\bigg\}.
\label{eq:U_0}
\end{equation}
The vertically-averaged velocity is therefore
\begin{equation}
\langle U_0\rangle = \tfrac{1}{2}H\frac{\partial\Sigma}{\partial X}-\tfrac{1}{3}H^2\bigg\{
\frac{\partial}{\partial X}\left(-\frac{1}{C}\frac{\partial^2 H}{\partial X^2}+\Phi_{\mathrm{surf}}\right)+\frac{r}{H}\frac{\partial}{\partial
X}\left[H\left(\frac{\partial c_0}{\partial X}\right)^2\right]
\bigg\},
\label{eq:uav}
\end{equation}
where we used the standard Laplace--Young free-surface boundary
condition to eliminate the pressure, and
\begin{equation}
C = \frac{\nu\rho D}{h_0\sigma_0\Small^2}=O\left(1\right).
\label{eq:C}
\end{equation}
Finally, by integrating the continuity equation in the $Z$-direction, we
obtain, in a standard manner, an equation for free-surface variations,
\begin{equation}
\frac{\partial H}{\partial T}+\frac{\partial}{\partial X}\left(H\langle U_0\rangle\right)=0.
\label{eq:h_eqn_ornament}
\end{equation}

The inclusion of the surface-tension terms requires further elucidation.  In the long-wave limit, the normal-stress condition reduces to $p|_h=-\sigma h_{xx}$, which in dimensionless form is
\[
P=-\frac{\sigma_0\Small^2h_0}{\nu \rho D}\left[1+\Small^qf\left(x\right)\right]H_{XX}.
\]
By promoting the constant $C={\nu\rho D}/\left(h_0\sigma_0\Small^2\right)$ to $O\left(1\right)$, and by taking $q>1$, this equation reduces to $P=-C^{-1}H_{xx}$, as in Eq.~\eqref{eq:uav}.  Similarly, the normal-stress condition reduces to $\nu\rho\left(\partial u/\partial z\right)_h=\partial\sigma/\partial x$, which in dimensionless form reads
\[
\frac{\partial U}{\partial Z}=\frac{\sigma_0 h_0\Small^q}{\nu\rho D}\frac{\partial f}{\partial X}.
\]
Taking $q=2$ gives a contribution to the shear-stress balance at lowest order, $\partial U/\partial Z=\partial \Sigma/\partial X$, $\Sigma=f\left(X\right)/C$, as in Eq.~\eqref{eq:sigma_x}.  Going to higher exponents $q>2$ suppresses this contribution.

Let us assemble our results, restoring the lower-case fonts and omitting
ornamentation over the constants.  The height equation~\eqref{eq:h_eqn_ornament}
becomes
\begin{subequations}
%\begin{gather}
\begin{equation}
\frac{\partial h}{\partial t}+\frac{\partial J}{\partial x}=0,
\end{equation}
while the concentration equation~\eqref{eq:conc_eqn_ornament} becomes
\begin{equation}
\frac{\partial}{\partial t}\left(c h\right)+\frac{\partial}{\partial x}\left(Jc\right)=\frac{\partial}{\partial{x}}\left(h\frac{\partial\mu}{\partial{x}}\right),
\end{equation}
where
\begin{equation}
J=\tfrac{1}{2}h^2\frac{\partial\sigma}{\partial{x}}-\tfrac{1}{3}h^3\bigg\{\frac{\partial}{\partial{x}}\left(-\frac{1}{C}\frac{\partial^2{h}}{\partial{x}^2}
+\phi\right)+\frac{r}{h}\frac{\partial}{\partial{x}}\left[h\left(\frac{\partial{c}}{\partial{x}}\right)^2\right]\bigg\},
\end{equation}
and
\begin{equation}
\mu=c^3-c-\Cn^2\frac{1}{h}\frac{\partial}{\partial{x}}\left(h\frac{\partial{c}}{\partial{x}}\right),
\end{equation}%
\label{eq:model}%
\end{subequations}%
and where we have the nondimensional constants
\begin{equation}
r=\frac{\Small^2\beta\gamma}{D\nu},\qquad \Cn=\frac{\Small\sqrt{\gamma}}{h_0},\qquad
C=\frac{\nu\rho D}{h_0\sigma_0\Small^2}.
\end{equation}
The boundary conditions are inherited from the full Navier--Stokes Cahn--Hilliard equations:  Since $J$ contains a depth-averaged velocity, we write it as $J:=hu$.  Thus, the concentration $c$, the chemical potential $\mu$, and the flux $J$ are either periodic functions in the lateral direction, or satisfy the no-flux conditions $c_x=\mu_x=J=0$ on the lateral boundaries.
Equations~\eqref{eq:model}, together with the boundary conditions described, are the thin-film NSCH equations.  The integral quantities defined in Eq.~\eqref{eq:integral_quantities} are manifestly conserved,
while the free surface and concentration are coupled.  The term $\left(h^2/2\right)\sigma_x$ in the flux $J$ represents a driving force, which can be externally prescribed, or a function of the concentration $c$.  In either case, the inclusion of this term can have a substantial effect on the behaviour of the system.    For the rest of this study, this term is set to zero; its inclusion is discussed elsewhere by the present authors~\cite{ONaraigh2007}, and by others~\cite{Oron2008}.

In view of the severe constraint $\Cn=\delta\sqrt{\gamma}/h_0=O\left(1\right)$, some discussion about the applicability of Eqs.~\eqref{eq:model} to real systems is warranted.  This constraint is the condition that the mean thickness of the film be much smaller than the transition-layer thickness.  In experiments involving the smallest
film thicknesses attainable ($10^{-8}$ m)~\cite{Sung1996}, this
condition is naturally satisfied.  
Furthermore, in certain situations far from this limiting case, variations in the domain structure in the vertical direction are suppressed, and a system of equations with no vertical ($z$-) dependence, such as Eqs.~\eqref{eq:model}, is appropriate.  This kind of situation arises when
external effects such as the air-fluid and
fluid-substrate interactions do not prefer one binary fluid component
or another; hence, the dimensionality of the film is reduced, and the balance laws implied by Eqs.~\eqref{eq:model} are applicable.

\section{The model equations: analysis}
\label{sec:analysis}

The choice of potential $\phi_0$ determines the behaviour of solutions.
In this section, we perform a linear-stability analysis on the model equations~\eqref{eq:model} and identify the pattern-formation mechanism.  We also develop results for the non-linear regime using the theory of \textit{a priori} bounds.  These are bounds on norms of the solution $\left(h,c\right)$ that are obtained without assuming any prior knowledge of the solution.  Throughout this section, the driving force resulting from surface-tension gradients is set to zero.  

The first step in our analytical study is to find the circumstances under which the constant state $\left(h_0,c_0\right)$ is unstable to a small-amplitude, initial perturbation $\left(\delta h_0,\delta c_0\right)$.  This perturbation evolves in time to a state $\left(\delta h,\delta c\right)\left(x,t\right)$, which satisfies the linearized version of equations~\eqref{eq:model}.  By writing down a wave ansatz $\left(\delta h,\delta c\right)\propto e^{ikx}$, we obtain an eigenvalue equation,
\[
\frac{d}{dt}\left(\begin{array}{c}\delta h\\\delta c\end{array}\right)=
\left(\begin{array}{cc}\frac{h_0^3}{3}\left[-k^4-k^2\phi_0'\left(h_0\right)\right]&0\\
0&-k^2\left(3c_0^2-1\right)-\Cn^2k^4\end{array}\right)
\left(\begin{array}{c}\delta h\\\delta c\end{array}\right),
\]
with eigenvalues
\begin{equation}
\begin{split}
\lambda_h &= -\frac{h_0^3 k^2}{3} \left[\frac{k^2}{C}+\phi_0'\left(h_0\right)\right],\\
\lambda_c &= -\left(3 c_0^2-1\right)k^2 - \Cn^2 k^4.
\end{split}
\label{eq:sigma}
\end{equation}
Thus, there are two routes to instability.  The system can become unstable as a result of substrate-film interactions if $\phi_0'\left(h_0\right)<0$.  Such an interaction will often lead to rupture~\cite{Oron1997}.    If this route is suppressed, then film rupture may be prevented, but the second route to instability is also relevant.  This is accessible when $c_0$ is in the spinodal range $ \left|c_0\right|<1/\sqrt{3}$.
Thus, even when the first route to instability is not accessible, a critical mixed state will phase separate in a manner similar
 to the classical Cahn--Hilliard fluid, as described in Sec.~\ref{sec:intro}.

While the linear analysis is helpful to describe early-stage evolution, it sheds no light on the behaviour at later times.  We therefore turn to the non-linear analysis of the problem~\eqref{eq:model}.
  The non-linear analysis centres on finding bounds for a given solution $\left(h,c\right)$.  To do this, it is necessary to construct a Lyapunov functional, that is, a non-negative, non-increasing functional of the solution $\left(h,c\right)$.  It is certainly the case that we can find a non-increasing functional based on the solution pair $\left(h,c\right)$, which is a kind of energy for the problem:
\begin{prop}[Existence of a decreasing functional]
\label{prop:lyapunov}
Given a smooth solution $\left(h,c\right)$ to the equations~\eqref{eq:model}, positive in the sense that $h\left(x,t\right)>0$, and a continuous potential function $\phi_0$, then the functional
\begin{equation}
\mathcal{F}\left[h,c\right]=\int_0^L{dx}\, \left[\frac{1}{2C}\left(\frac{\partial{h}}{\partial{x}}\right)^2+\int^h\phi_0\left(s\right)ds\right]
+ \frac{r}{\Cn^2}\int_0^L{dx}\,  h\left[\tfrac{1}{4}\left(c^2-1\right)^2+\frac{\Cn^2}{2}\left(\frac{\partial{c}}{\partial{x}}\right)^2\right]
\label{eq:Fhc}
\end{equation}  
is non-increasing, $\dot{\mathcal{F}}\leq 0$.
\end{prop}
The proof of this claim is readily obtained by a straightforward time-differentiation of $\mathcal{F}\left[h,c\right]$ and application of the equations~\eqref{eq:model}, together with the no-flux or periodic boundary conditions.  We find,
\begin{equation}
\dot{\mathcal{F}}=-\int_0^L dx \left(\frac{3J^2}{h^3}+h\mu_x^2\right)\leq 0.
\label{eq:Fhc_min}
\end{equation}
We build on this result by focussing on the following class of potential whose anti-derivative is positive:
\begin{equation}
\phi_1\left(s\right):=-\int_s^a\phi_0\left(s'\right)ds'>0,\qquad 0<s<a,
\label{eq:pos}
\end{equation}
where $a$ is an arbitrary reference height.  Using Prop.~\ref{prop:lyapunov} and the condition~\eqref{eq:pos}, we obtain a Lyapunov functional for the positive solution $\left(h,c\right)$, $h>0$:
\begin{prop}[Existence of a positive Lyapunov functional]
\label{prop:lyapunov2}
For a smooth solution $\left(h,c\right)$ to the equations~\eqref{eq:model}, positive in the sense that $h\left(x,t\right)>0$, and for a potential function $\phi_0$ with positive anti-derivative, there is an associated Lyapunov functional.
\end{prop}
To verify this claim, it suffices to note that since $\phi_1>0$ for the class of potential functions under consideration, all terms in the functional $\mathcal{F}\left[h,c\right]$ (Eq.~\eqref{eq:Fhc}) are positive, and thus $\mathcal{F}$ is a positive, non-increasing function of time, i.e. a Lyapunov functional.

The boundedness result $\mathcal{F}\left(t\right)\leq \mathcal{F}\left(0\right)$ provides a regularity condition on the height $h\left(x,t\right)$, although this is valid only in a single spatial dimension.  
\begin{prop}[H\"older continuity of $h\left(x,\cdot\right)$]
\label{prop:holder}
If $\left(h,c\right)$ is a smooth, positive solution to the equations~\eqref{eq:model}, in the sense that $h\left(x,t\right)>0$, and if the potential function $\phi_0$ has a positive anti-derivative, then $h\left(x,\cdot\right)$ is H\"older continuous, with time-independent H\"older constant $k_H$.
\end{prop}
Proposition~\ref{prop:holder} follows from the \textit{a priori} bounds
\begin{equation}
\frac{1}{2C}\int_0^L dx\, h_x^2\leq \mathcal{F}\left(t\right)\leq \mathcal{F}\left(0\right),
\end{equation}
and from the use of H\"older's inequality on the following string of relations:
\[
 |h\left(x_2\right)-h\left(x_1\right)|=\left|\int_{x_1}^{x_2}h_xdx\right|
 \leq\int_{x_1}^{x_2}dx\,|h_x|\\
 \leq|x_2-x_1|^{1/2}\sqrt{\int_0^L dx\,h_x^2}
 \leq k_H|x_2-x_1|^{1/2},
 \]
where $k_H=\sqrt{2C\mathcal{F}\left(0\right)}$ is the time-independent H\"older constant. 
As an immediate corollary of this result, we obtain an upper bound on the height field:
\begin{prop}[An upper bound on the height field]
\label{prop:h_upper}
If $\left(h,c\right)$ is a smooth, positive solution to the equations~\eqref{eq:model}, in the sense that $h\left(x,t\right)>0$, and if the potential function $\phi_0$ has a positive anti-derivative, then $h\left(x,\cdot\right)$ is bounded above.
\end{prop} 
Since the free energy contains a term in the $L^2$-norm of $h^{1/2}c_x$, a similar result exists for the concentration field, provided $h>0$ everywhere.   Loss of control over the minimum value of $h$ therefore implies loss of control over the concentration gradient.  This suggests that blowup of gradients and film rupture are related, a claim which we demonstrate numerically in Sec.~\ref{sec:numerics_rupture}.
Such extreme events are avoided when a repulsive film-substrate interaction is present, in which case a positive lower bound on $h$ exists; it is to that result that we now turn.

Given the form of Eqs.~\eqref{eq:model}, regularity of a given solution $\left(h,c\right)$ is guaranteed only when a lower bound on $h$ is obtained, in addition to the upper bounds just provided.  To derive such a result, we first specialize to the potential
\begin{equation}
\phi_0=-\frac{G}{2s^3},\qquad G>0.
\label{eq:phidef0}
\end{equation}
from which a more general result will follow.
\begin{prop}[No-rupture condition for the potential in Eq.~\eqref{eq:phidef0}]
\label{prop:norupture_h3}
If $\left(h,c\right)$ is a smooth, positive solution to the equations~\eqref{eq:model}, in the sense that $h\left(x,t\right)>0$, and if the potential function $\phi_0$ has the form given by Eq.~\eqref{eq:phidef0}, then there is an \textit{a priori}, time-independent lower bound on $h$.
\end{prop} 
Note first of all that the potential~\eqref{eq:phidef0} has a positive anti-derivative, $\phi_1\left(s\right)=Gs^{-2}$, where the reference height $a$ in Eq.~\eqref{eq:pos} is set to $a=\infty$.  Thus, there is a Lyapunov functional for the solution $\left(h,c\right)$, and hence,
\begin{equation}
\int_0^L dx\, Gh^{-2}=\int_0^L dx\, \phi_1\left(h\right)\leq \mathcal{F}\left(t\right)\leq \mathcal{F}\left(0\right).
\label{eq:Gbd}
\end{equation}
Using the H\"older continuity of $h$, 
\begin{equation}
h^2\leq h_{\mathrm{min}}^2+2h_{\mathrm{min}}k_HL^{1/2}+k_H^2|x-x_\mathrm{min}|
 \leq h_{\mathrm{min}}^2+2h_{\mathrm{min}}k_HL^{1/2}+k_H^2 x,
\label{eq:holderh}
\end{equation}
where $k_H=\sqrt{2C\mathcal{F}\left(0\right)}$ is the H\"older constant for $h$.  Using results~\eqref{eq:Gbd} and~\eqref{eq:holderh}, 
\begin{equation}
G\int_0^L\frac{dx}{h_\mathrm{min}^2+2h_\mathrm{min}k_HL^{1/2}+k_H^2 x}\leq \mathcal{F}\left(0\right).
\label{eq:inth}
\end{equation}
By integrating this equation, we arrive at the relation
\begin{equation}
\log\left(1+\frac{ k_H^2 L}{h_\mathrm{min}^2+2k_H L^{1/2}h_\mathrm{min}}\right)\leq \frac{\mathcal{F}\left(0\right)k_H^2}{G}.
\label{eq:phi_ineq}
\end{equation}
Let us examine the properties of the function $\log\left[1+k_H^2L/\left(s^2+2k_H L^{1/2}s\right)\right]$.  It tends to infinity as $s\rightarrow0$, and tends to zero as $s\rightarrow\infty$.  It is also monotone-decreasing over $s\in\left(0,\infty\right)$.  Thus, the equation $\log\left[1+k_H^2L/\left(s^2+2k_H L^{1/2}s\right)\right]=\mathcal{F}\left(0\right)k_H^2/G$ has precisely one positive root for $\mathcal{F}\left(0\right)\neq0$, which we call $s_*$.  To satisfy the inequality~\eqref{eq:phi_ineq}, it must be the case that
\[
h_\mathrm{min}\geq s_*>0.
\] 
For the potential $\phi_0=-\left(G/2\right)s^{-3}$ (Eq.~\eqref{eq:phidef0}), the root $s_*$ can be obtained explicitly:
\[
h_\mathrm{min}\geq s_*=k_H L^{1/2}\left[\sqrt{\frac{1}{\exp\left(\mathcal{F}\left(0\right)k_H^2/G\right)-1}}-1\right]>0.
\] 
This completes the proof of the no-rupture condition for the potential~\eqref{eq:phidef0}.  
 
To arrive at a no-rupture condition for a general potential, we introduce the function
\begin{equation}
\phi_2\left(s\right):=\int_0^L dx\, \phi_1\left(s+k_H x^{1/2}\right).
\label{eq:PHIdef}
\end{equation}
Based on Prop.~\ref{prop:norupture_h3}, we write down sufficient conditions on $\phi_1$ and $\phi_2$ for the existence of a positive lower bound on $h$:
\begin{prop}[A sufficient condition to avoid rupture]
\label{prop:norupture_gl}
If $\left(h,c\right)$ is a smooth, positive solution to the equations~\eqref{eq:model}, in the sense that $h\left(x,t\right)>0$, if the anti-derivative of the potential function $\phi_0$ is a positive, non-increasing function, and moreover, if $\phi_2$ satisfies the conditions
 \begin{eqnarray}
 \lim_{s\rightarrow0}\phi_2\left(s\right)&=&\infty,\nonumber\\
 \lim_{s\rightarrow\infty}\phi_2\left(s\right)&\leq&0,
 \label{eq:PHIcond}
\end{eqnarray}
then a positive \textit{a priori} lower bound on $h\left(x,t\right)$ exists, independent of time.
\end{prop}
The proof of Prop.~\ref{prop:norupture_gl} is in the same spirit as that of Prop.~\ref{prop:norupture_h3}.  Given the positivity of the anti-derivative $\phi_1$, a Lyapunov exponent exists (Prop.~\ref{prop:lyapunov}), and thus we have the bound $\int dx\, \phi_1\left(h\right)\leq \mathcal{F}\left(0\right)$.  Using the H\"older continuity of $h$ (Prop.~\ref{prop:holder}), and the condition that the $\phi_1$ should be a non-increasing function,
\begin{eqnarray*}
h\left(x,t\right)&\leq& h_\mathrm{min}+k_H x^{1/2},\\
\phi_1\left(h\left(x,t\right)\right)&\geq& \phi_1\left(h_\mathrm{min}+k_H x^{1/2}\right).
\end{eqnarray*}
Using the bound $\int dx\, \phi_1\left(h\right)\leq \mathcal{F}\left(0\right)$,
\[
\phi_2\left(h_\mathrm{min}\right)=\int_0^L dx\, \phi_1\left({h_\mathrm{min}+k_H x^{1/2}}\right)
\leq \mathcal{F}\left(0\right).
\]
Given the conditions~\eqref{eq:PHIcond}, there exists at least one solution to the equation $\phi_2\left(s\right)=\mathcal{F}\left(0\right)$, for $\mathcal{F}\left(0\right)>0$.  Indeed, since $\phi_1$ is non-increasing, so too is $\phi_2$, and thus this equation has precisely one solution, which we call $s_*$.
Then, for the condition $\phi_2\left(h_\mathrm{min}\right)\geq \mathcal{F}\left(0\right)$ to be satisfied on the interval $\left(0,\infty\right)$, we must take
\[
h_\mathrm{min}\geq s_{*}>0.
\]
Finally, using this theory, we investigate the potential
\begin{equation}
\phi_0=-\frac{G}{n s^{n+1}},
\label{eq:phidef1}
\end{equation}
and calculate the $n$-values for which a no-rupture condition can be found.
\begin{prop}[Conditions on the potential~\eqref{eq:phidef1} to avoid rupture]
\label{prop:norupture_n}
If $\left(h,c\right)$ is a smooth, positive solution to the equations~\eqref{eq:model}, in the sense that $h\left(x,t\right)>0$, and if the potential function $\phi_0$ is given by Eq.~\eqref{eq:phidef1}, then a no-rupture condition is guaranteed to hold for $n\geq2$.
\end{prop}
The proof of Prop.~\ref{prop:norupture_n} follows by a straightforward evaluation of the integral~\eqref{eq:PHIdef}.  The case $n=2$ is covered by Prop.~\ref{prop:norupture_h3}.  Thus, we focus on the case $n\neq2$, where the integral $\phi_2\left(s\right)$ has the value
\begin{equation}
\phi_2\left(s\right)=\frac{2G}{\left(n-2\right)\left(n-1\right)k_H^n}\frac{1}{\alpha^{n-2}}\left[
1-\left(\frac{\alpha}{\alpha+1}\right)^{n-1}\left(1+\frac{n-1}{\alpha}\right)
\right],
\qquad \alpha=s/k_H,
\label{eq:PHIpower}
\end{equation}
and where we have set $L=1$.  For $n>2$ the conditions~\eqref{eq:PHIcond} hold: $\lim_{s\rightarrow0}\phi_2\left(s\right)=\infty$, $\lim_{s\rightarrow\infty}\phi_2\left(s\right)=0$; note also that $\phi_2\left(s\right)$ is non-increasing.  Thus, the equation $\phi_2\left(s\right)=\mathcal{F}\left(0\right)$ has exactly one positive root $s_*$, and this serves as a lower bound on $h$, $h_\mathrm{min}\geq s_*>0$.  Note
\begin{figure}[htb]
\includegraphics[width=.5\textwidth]{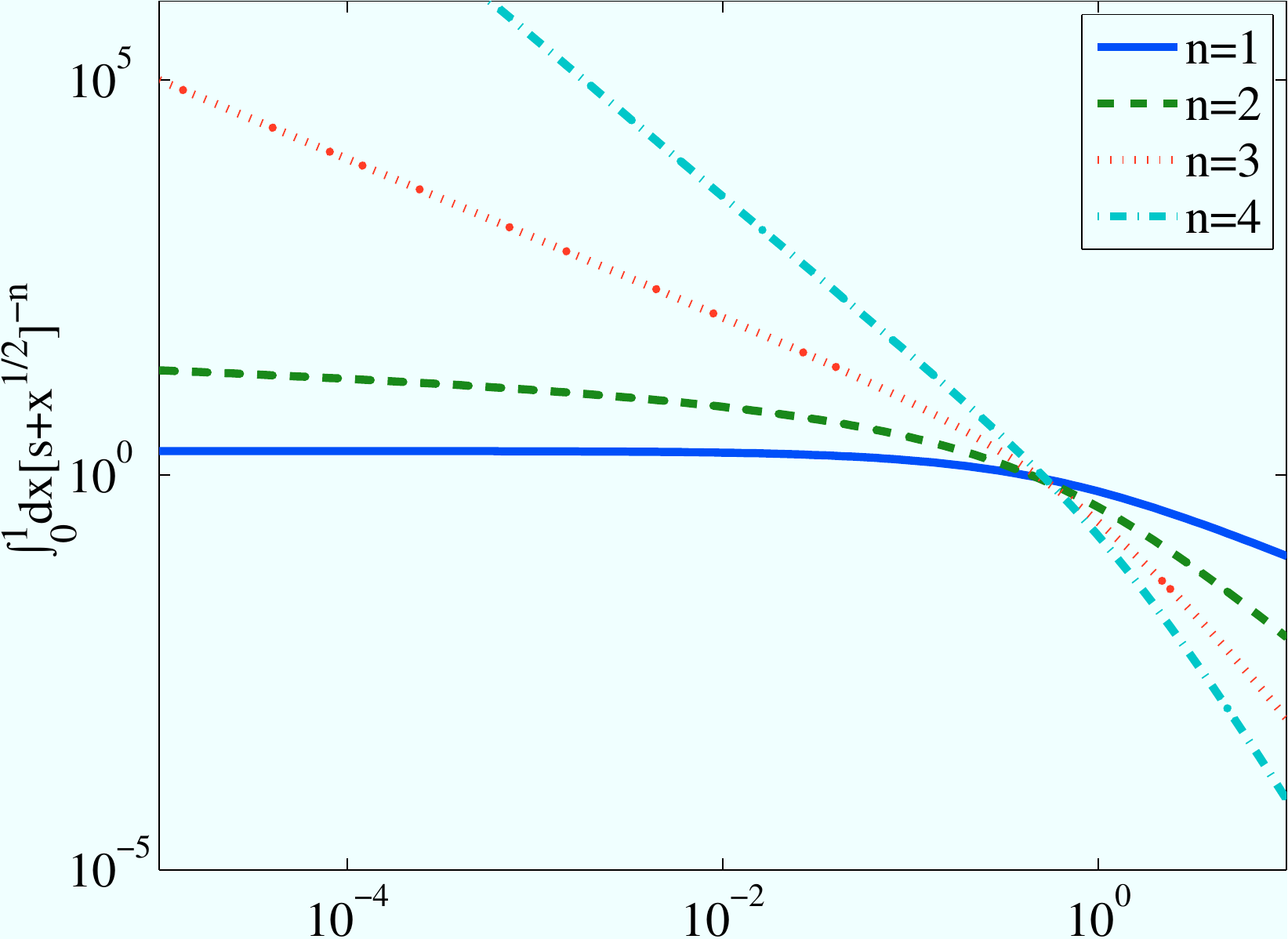}
\caption{
A plot of the integral $\int_0^L dx\left(s+x^{1/2}\right)^{-n}$ as a function of $s$.  For $n\geq2$ the integral diverges as $s\rightarrow0$ and tends to zero as $s\rightarrow\infty$, allowing for the existence of a positive root for the equation $\int_0^L dx\left(s+x^{1/2}\right)^{-n}=\mathrm{const.}>0$.
}
\label{fig:integral}
\end{figure}
that the relation given in Eq.~\eqref{eq:PHIpower} fails to satisfy conditions~\eqref{eq:PHIcond} when $n<2$.  Thus, a sufficient condition for the solution $\left(h,c\right)$ to possess a no-rupture condition is for the potential $\phi_0$ to have the form given in Eq.~\eqref{eq:phidef1}, with $n\geq2$.  
This analysis is also described schematically in Fig.~\ref{fig:integral}, where a plot of the integral $\int_0^L dx\left(s+x^{1/2}\right)^{-n}$ is shown as a function of $s$.  For $n\geq2$ the integral diverges as $s\rightarrow0$ and tends to zero as $s\rightarrow\infty$, which provides for the existence of a positive root for the equation $\int_0^L dx\left(s+x^{1/2}\right)^{-n}=\mathrm{const.}>0$.

Two questions arise from our results.  The first involves the form of the potential used in the derivation of a no-rupture condition: can we replace the repulsive power-law form with a more general function and still obtain a no-rupture condition?  The answer to this question comes readily through the observation that our construction 
of a positive lower bound for $h$ relies only on the surface-tension and body-force components of the free energy, namely
\[
\mathcal{F}_1=\int_0^Ldx\left[\frac{1}{2C}h_x^2+\phi_1\left(h\right)\right].
\]
Thus, Propositions~\ref{prop:norupture_h3}--\ref{prop:norupture_n} can be viewed in the context of the PDE theory for a single variable (the free-surface height), where a no-rupture condition exists~\cite{Grun2001}, both for the power-law type potential considered here and for the Lennard--Jones potential
$\phi_0=-Gs^{-n_1}+Bs^{-n_2}$, where $G$ and $B$ are positive constants and $n_1>\max\left(n_2,\min\left(1+2n_2,3\right)\right)$~\cite{Grun2001}.
Hence, our regularity results are generalizable to this wider class of potential.

Having weakened the sufficient condition for the avoidance of rupture, it is reasonable to ask, is the presence of a suitable potential even a necessary condition?
This question is motivated by the single-variable theory for the free-surface height, where under certain conditions an entropy functional facilitates the construction of an \textit{a priori} lower bound on the height $h\left(x,t\right)$~\cite{Friedman1990}.
The entropy is obtained through the following steps, which we describe for the single-variable case $3h_t+\{h^n\left[\left(h_{xxx}/C\right)-\phi_x\right]\}_x=0$:
\begin{enumerate}
\item Identify the power of $h$ that is a factor in the flux $J$; this is the \textit{mobility}, $m_0$.  For the single-variable case, $m_0\left(h\right)=h^n$.
\item Obtain the function $m_1\left(s\right)=\int^s ds' \int^{s'}ds''\left[m\left(s''\right)\right]^{-1}$.
\item The entropy is then defined as $\mathcal{S}=\int_0^L dx\,
  m_1\left(h\right)$; for the single-variable case, this is
  $\mathcal{S}=\int_0^L dx\, h^{-n+2}$ (we have omitted the
  unimportant constant of proportionality).
\end{enumerate}
For the single-variable case,
\begin{equation}
\mathcal{S}\left(t\right)+\frac{2}{3C}\int_0^t dt'\int_0^L dx\, h_{xx}^2+\tfrac{2}{3}\int_0^t dt'\int_0^L dx\,h_x^2\phi_0'\left(h\right)=\mathcal{S}\left(0\right)>0,
\label{eq:entropy}
\end{equation}
and the entropy is a non-increasing functional of $t$, $d \mathcal{S}/dt\leq0$.
Setting $\phi_0=0$ in Eq.~\eqref{eq:entropy} gives the relation
\begin{equation}
\int_0^L \frac{dx}{\left[h\left(x,t\right)\right]^{n-2}}+\frac{2}{3C}\int_0^t dt'\int_0^L dx\, h_{xx}^2=\mathcal{S}\left(0\right)>0.
\label{eq:entropy1}
\end{equation}
For $n\geq4$, H\"older continuity combined with the bound in Eq.~\eqref{eq:entropy1} enables the construction of a pointwise lower bound on $h$.  When $n=3$ (the case considered in this paper), no such pointwise bound exists; then, estimates for the entropy based on the H\"older continuity of $h$ are non-singular in the minimum height.  However, a more definitive obstacle than this exists when we seek to construct an entropy functional for the system~\eqref{eq:model}, namely, that the entropy functional implied by Eqs.~\eqref{eq:model} fails to be a non-increasing function of time:
\begin{prop}{\bf{(The time-derivative of the entropy associated with Eqs.~\eqref{eq:model} is not sign definite)}}  Given a smooth, positive solution  $\left(h,c\right)$ to the equations~\eqref{eq:model}, in the sense that $h\left(x,t\right)>0$, the rate of change of the functional $\int_0^Ldx\,h^{\alpha}$, $\alpha<0$, is not sign-definite.
\label{prop:no_entropy}
\end{prop}
This result is established by direct computation:
\begin{multline*}
\frac{d}{dt}\int_0^L dx\,h^{\alpha}=
-\frac{\alpha\left(\alpha-1\right)}{3C}\int_0^L dx\,\left(h^{\alpha+1}h_x\right)_x{h_{xx}}
\\
-\frac{\alpha\left(\alpha-1\right)}{3}\int_0^L dx\, h^{\alpha+1}h_x^2\phi_0'\left(h\right)
-\frac{r\alpha\left(\alpha-1\right)}{3}\int_0^L dx\,\left(h^\alpha h_x\right)_x h c_x^2.
\end{multline*}
No value of $\alpha$ can give the integral in the last term, namely
\[
\int_0^L dx\,\left(h^\alpha h_x\right)_x h c_x^2=
\int_0^L dx\left[\alpha h^{\alpha}c_x^2h_x^2+h^{\alpha+1}h_{xx}c_x^2\right]
\]
a definite sign.  Hence, the entropy functional fails to be non-increasing, and therefore cannot be used to construct a lower bound on $h$.  The existence of a suitable potential function in the evolution equation for $h$ is thus a necessary and sufficient condition for the avoidance of rupture.  Indeed, the existence of the backreaction in the equations~\eqref{eq:model}, and its consequences for the entropy functional, suggest that it promotes rupture while the regularizing potential inhibits it.  To examine this effect further, we turn to numerical simulations.

\section{Numerical results}
\label{sec:height_dip}

\noindent 
In this section we obtain numerical solutions of the equations~\eqref{eq:model}, with and without the regularizing potential.  Our numerical method is described and validated in Appendix~A.  Where present, the regularizing potential is assigned the form $\phi=-G/h^3$, $G>0$, which satisfies the no-rupture condition given in Sec.~\ref{sec:analysis}.  
In the case where the no-rupture condition is satisfied, we study solutions that tend towards an equilibrium.  We then characterize this equilibrium through the solution of a boundary-value problem.  The tendency towards equilibrium is in agreement with the predictions of Sec.~\ref{sec:analysis}, where a linearly unstable state evolves to reduce the energy functional~\eqref{eq:Fhc}, such that domains of concentration form, separated by transition zones, across which the height of the film decreases markedly, forming `valleys'.  For the case where the no-rupture condition is not automatically satisfied ($G=0$), we perform numerical studies that suggest that rupture does indeed occur, and coincides with the development of a finite-time singularity.
Throughout this section, the driving force resulting from surface-tension gradients is set to zero.

\subsection{Numerical studies with a regularizing Van der Waals potential}
\label{sec:num_vdw}

We perform numerical simulations of the full equations~\eqref{eq:model},
with initial data comprising a perturbation away from the unstable steady
state $\left(h,c\right)=\left(1,0\right)$:
\[
h\left(x,0\right)=1,\qquad c\left(x,0\right)=0.01\sin\left[5\left(2\pi/L\right)x\right].
\]
We fix the parameters $G=C=1$, and $\Cn=0.1$.  The $\Cn$-value is small enough such that the transition width of domains is much smaller than the size of the computational domain $L=2\pi$.  We vary the parameter $r$ between $0.1$ and $1$ to examine the effects of the backreaction strength.  The calculations are carried out on a periodic domain, with $N=256$ gridpoints and a timestep $\Delta t=10^{-4}$.    The early-stage growth in the concentration  is governed by the linear theory of Eq.~\eqref{eq:sigma}.
\begin{figure}[htb]
\centering
\subfigure[]{
\includegraphics[width=0.35\textwidth]{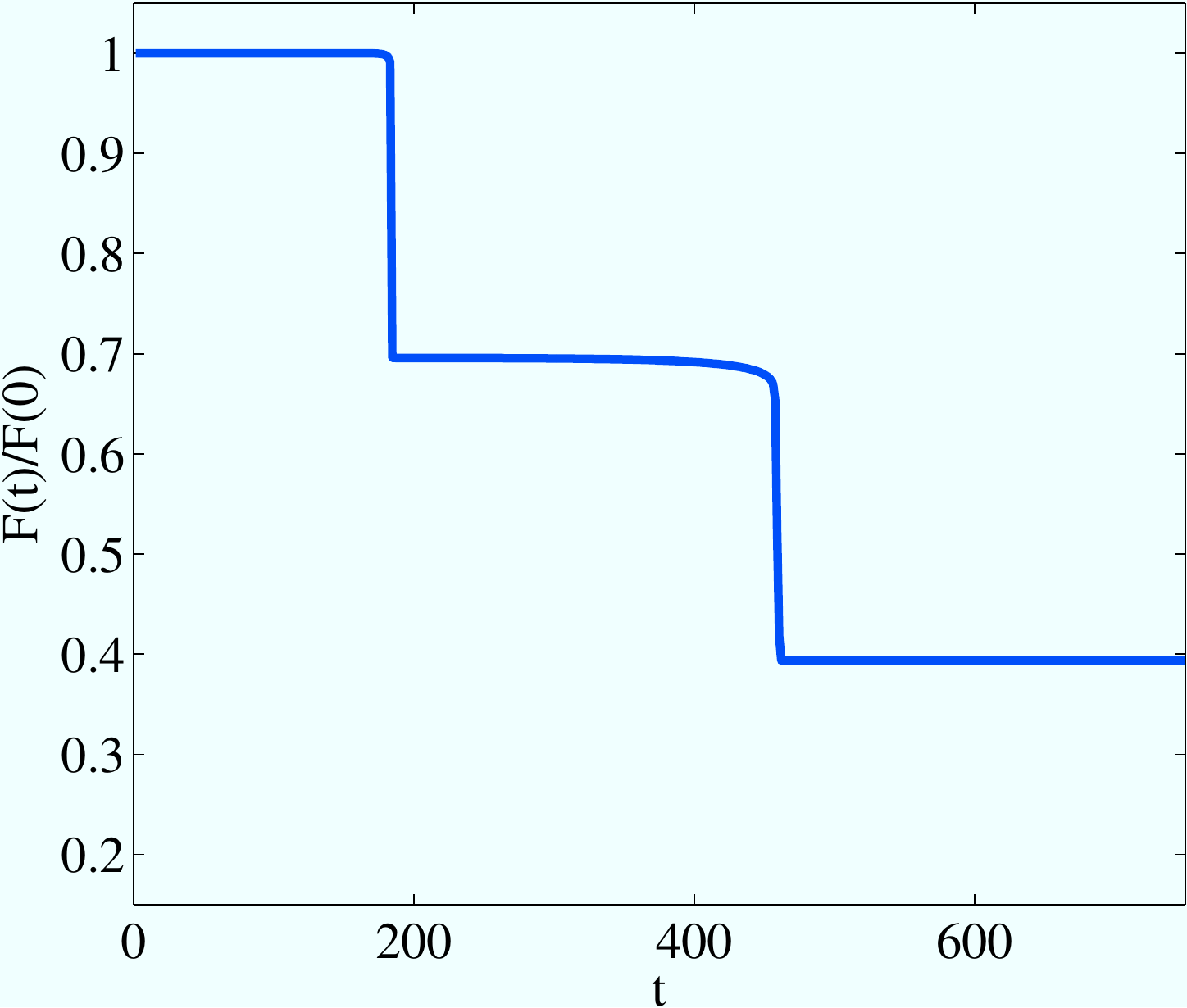}}
\subfigure[]{
\includegraphics[width=0.35\textwidth]{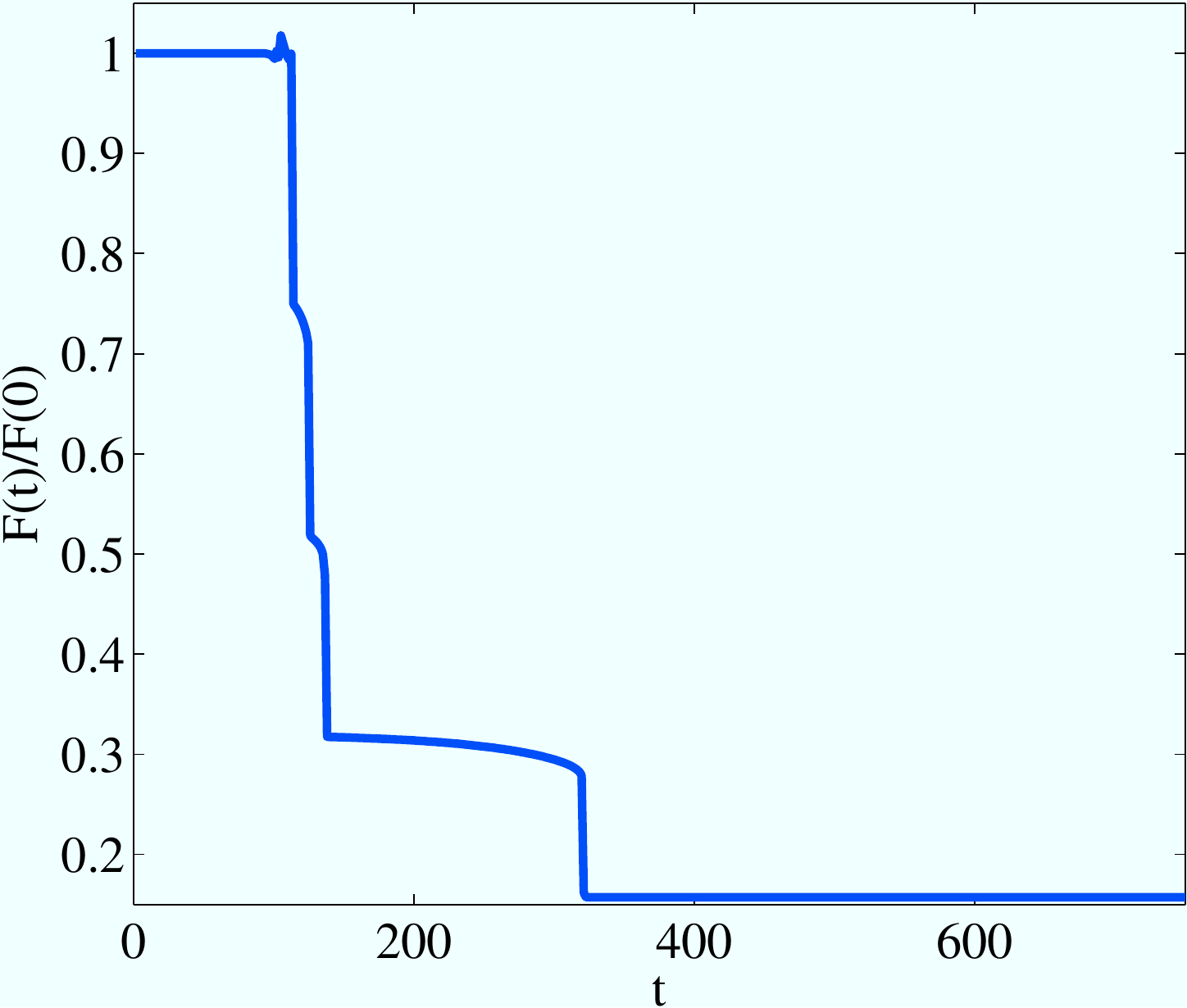}}
\caption{Numerical simulation of the decay of the free energy over time.  (a) $r=0.1$; (b) $r=1$ (the other parameters are kept constant: $C=G=1$, $\Cn=0.1$).  There is a very small, local increase in the free energy in case (b) due to numerical error.  This is small however, and is in contrast to the otherwise decreasing trend in the free energy.  The energy decreases particularly sharply when domains merge: the sharp drops correspond precisely to the domain-merging events in Fig.~\ref{fig:temporal1}.}
\label{fig:temporal}
\end{figure}
The free energy is a decreasing function of time (Fig.~\ref{fig:temporal}).  The average height is conserved exactly by the numerical simulation, while the mass $\int ch dx$ deviates 
from its initial value of zero within a range $\pm10^{-4}$ for the $r=0.1$ case, and $\pm10^{-2}$ for the $r=1$ case.
In the course of this evolution, the amplitude of the initial sinusoidal concentration field grows transiently in time; later on the positive and negative `domains' formed by each half-period of the sine function merge to form larger domains, each of larger amplitude.  At the same time, the free-surface height decreases in value at the borders of these domains, forming valleys.  Eventually, and as a consequence of the energy-minimization principle~\eqref{eq:Fhc_min},  only a single domain remains.  This evolution is best described visually, as in Fig.~\ref{fig:temporal1}, where spacetime plots of $c$ and $h$ show eventual coalescence into a pair of opposite-signed $c$-domains.  The domain coalescence happens more rapidly for the $r=1$ case, compared to the $r=0.1$ case.  This shows that the coupling of the free-surface height to the Cahn--Hilliard concentration far from arresting the domain coarsening, actually enhances it.
\begin{figure}[htb]
\centering
\subfigure[]{
\includegraphics[width=0.35\textwidth]{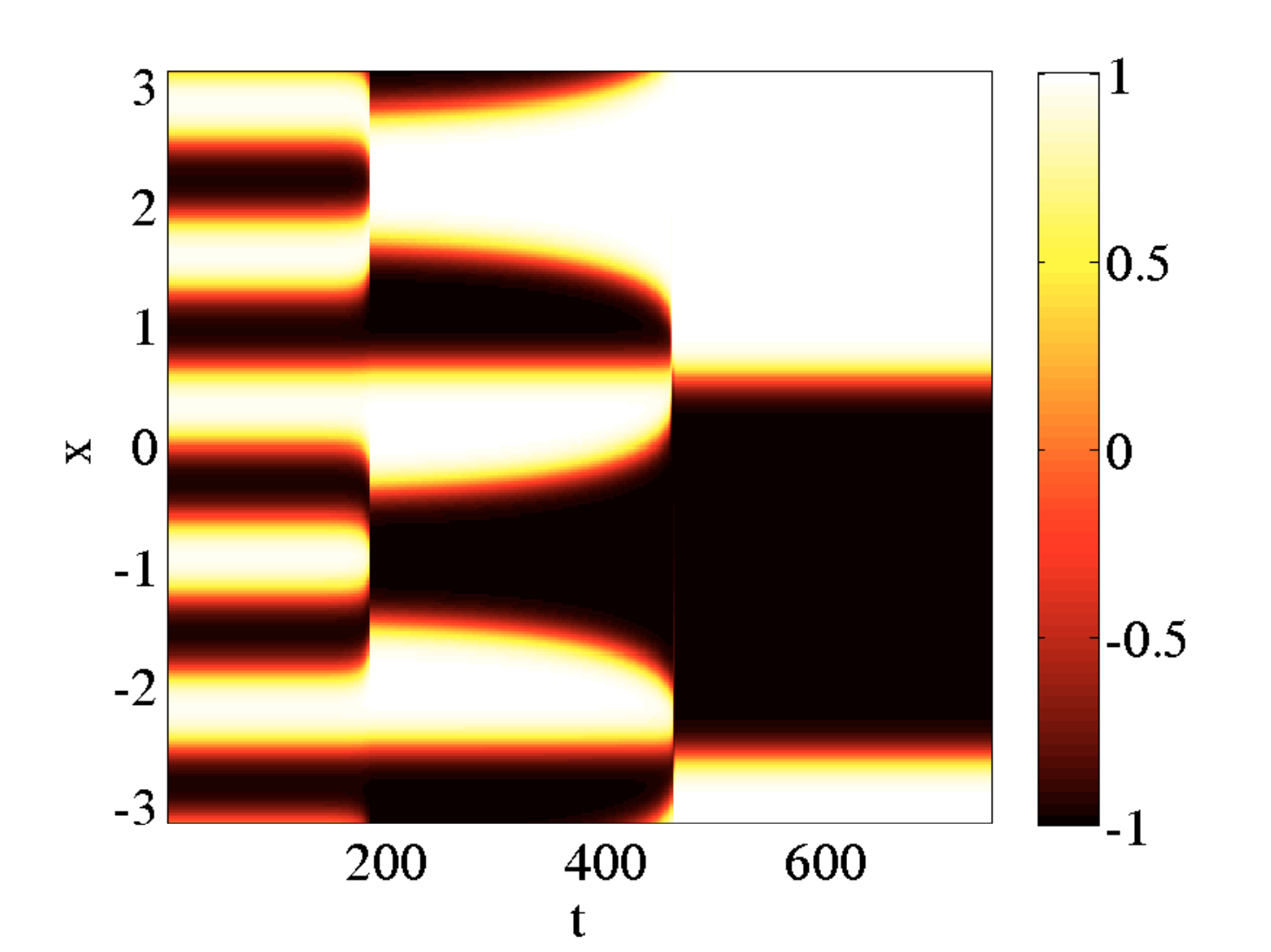}
}
\subfigure[]{
\includegraphics[width=0.35\textwidth]{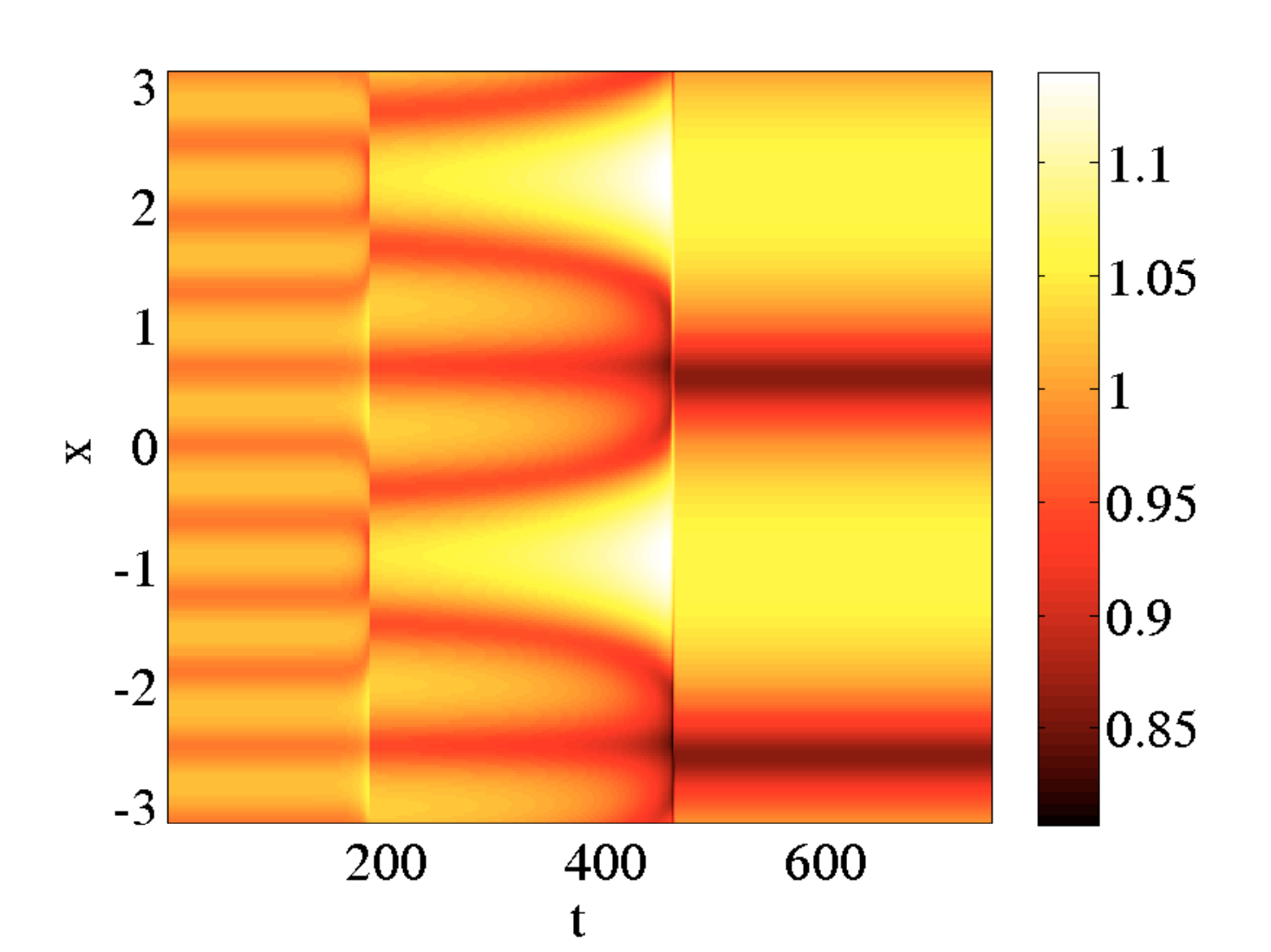}
}\\
\subfigure[]{
\includegraphics[width=0.35\textwidth]{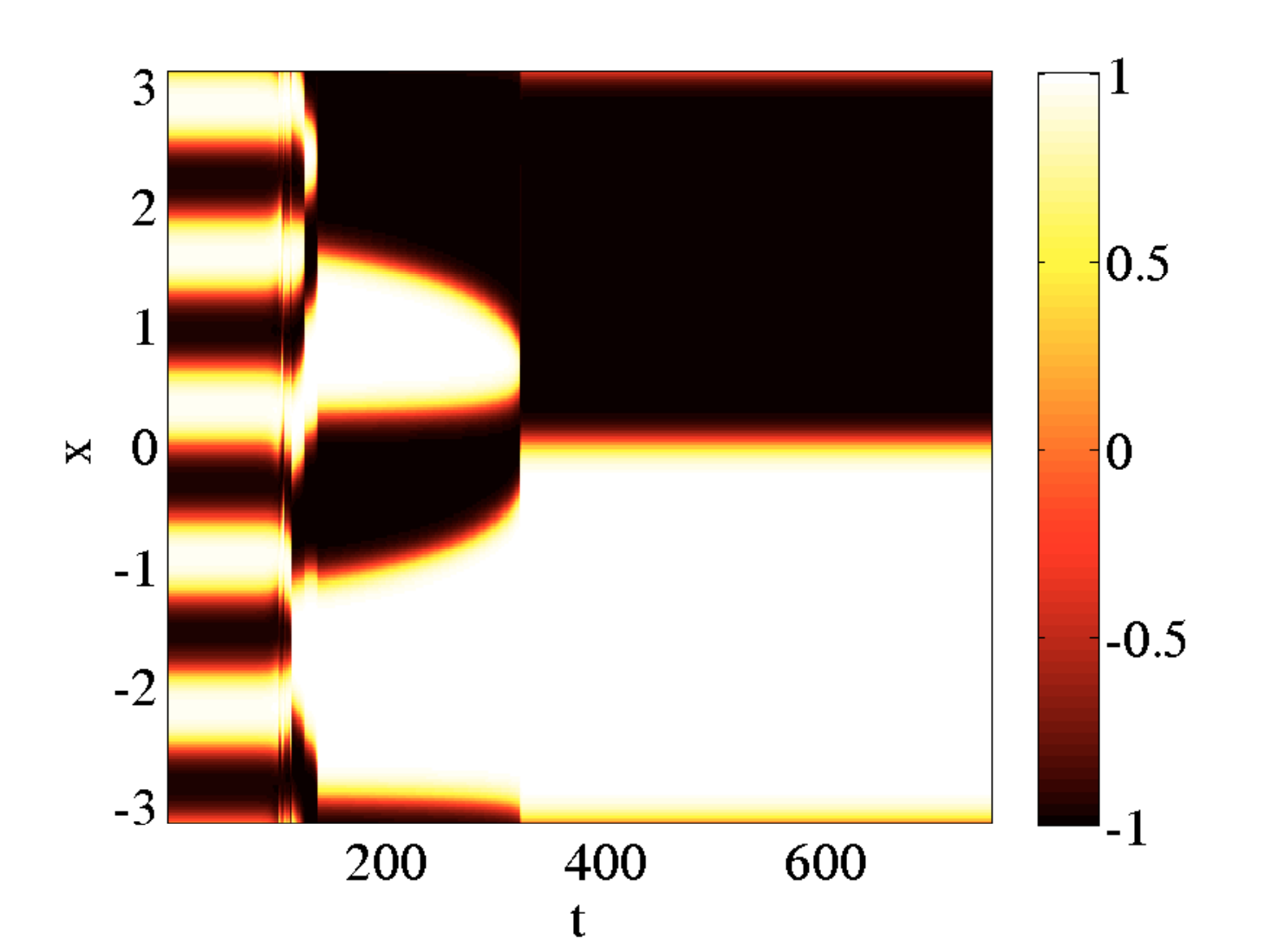}
}
\subfigure[]{
\includegraphics[width=0.35\textwidth]{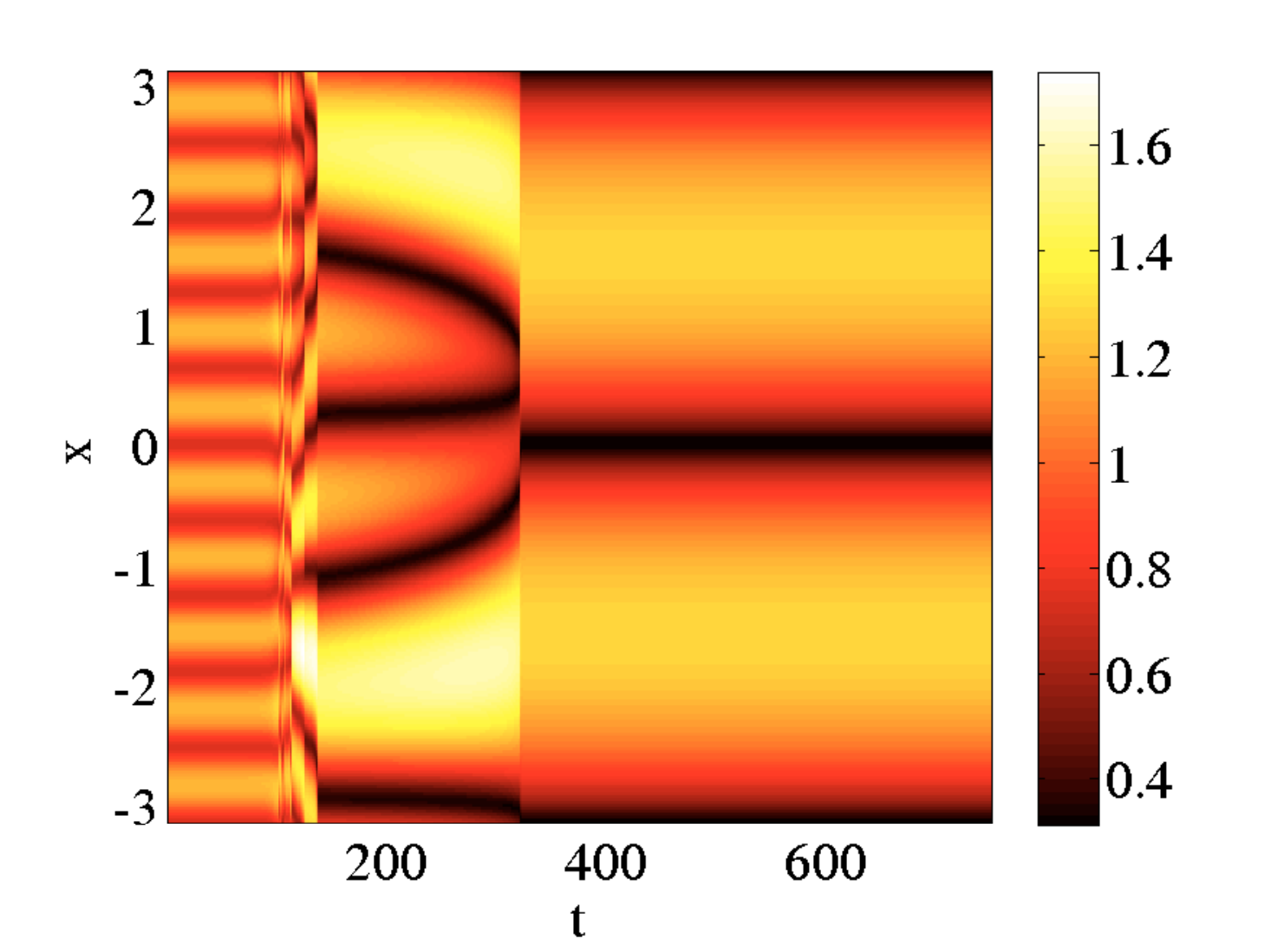}
}
\caption{Temporal evolution of the free-surface and concentration fields.  Across the top the backreaction strength is set to $r=0.1$.  Subfigure (a) shows the concentration for this case; (b) shows the free-surface height.  Across the bottom the backreaction strength has been increased to $r=1.0$.  Subfigure (c) shows the corresponding concentration; (d) shows the free-surface height.  The other parameters are kept constant: $C=G=1$, $\Cn=0.1$.  The domains coalesce until only a pair of opposite-signed domains remain.  The coalescence is more rapid for the $r=1$ case, compared to the $r=0.1$ case.
}
\label{fig:temporal1}
\end{figure}

The free surface and concentration evolve to an equilibrium state where the
salient feature is the formation of domains (intervals where $c\approx\pm1$)
that are separated by smooth transition zones, across which the free surface
dips below its average value.  We therefore shift focus to this state,
obtained by setting $\mu=\mathrm{constant}$, $u=0$ in Eq.~\eqref{eq:model}:
\begin{subequations}
\begin{equation}
\frac{1}{C}\frac{\partial^2 h}{\partial x^2}= \Cn^2 G \left(1-\frac{1}{h^3}\right)
+r\left[\tfrac{1}{4}\left(c^2-1\right)^2+\tfrac{1}{2}\left(\frac{\partial{c}}{\partial{x}}\right)^2\right],
\label{eq:eqm1}
\end{equation}
\begin{equation}
\frac{\partial^2 c}{\partial x^2}=c^3-c-\frac{1}{h}\frac{\partial h}{\partial
x}\frac{\partial c}{\partial x},
\end{equation}%
\label{eq:eqm}%
\end{subequations}%
where we have enforced the boundary conditions $h\left(\pm\infty\right)=1$,
$\mu\left(\pm\infty\right)=0$ and have rescaled lengths by $\Cn$.
For the case $C=\infty$ and $\rho\equiv r/\Cn^2G\ll1$, Eqs.~\eqref{eq:eqm}
have an asymptotic solution.  We find the $h$-equation
\begin{equation}
h=\biggl\{1+\rho\left[\tfrac{1}{4}\left(c^2-1\right)^2+\tfrac{1}{2}\left(\frac{\partial
c}{\partial x}\right)^2\right]\biggr\}^{-1/3}.
\label{eq:r_dep}
\end{equation}
Hence, the $c$-equation is 
\begin{equation}
\frac{\partial^2 c}{\partial x^2}=\frac{1+\frac{1}{4}\rho\left(c^2-1\right)^2+\frac{5}{6}\rho\left(\frac{\partial
c}{\partial x}\right)^2}
{1+\frac{1}{4}\rho\left(c^2-1\right)^2+\frac{1}{6}\rho\left(\frac{\partial
c}{\partial x}\right)^2}\left(c^3-c\right).
\label{eq:c_eqn_no_st}
\end{equation}
For small $\rho$, the solution is $c=\tanh\left(x/\sqrt{2}\right)+O\left(\rho\right)$
and hence
\begin{equation}
h=1-\tfrac{1}{3}\rho\,\mathrm{sech}^4\left(\frac{x}{\sqrt{2}}\right)+O\left(\rho^2\right),\qquad\rho\ll1.
\label{eq:h_large_r}
\end{equation}
Thus, in this limiting case, the height profile is approximately constant
($h=1$) except in the transition region of the concentration field, where it dips.

\begin{figure}[htb]
\centering
\subfigure[]{
\includegraphics[width=0.3\textwidth]{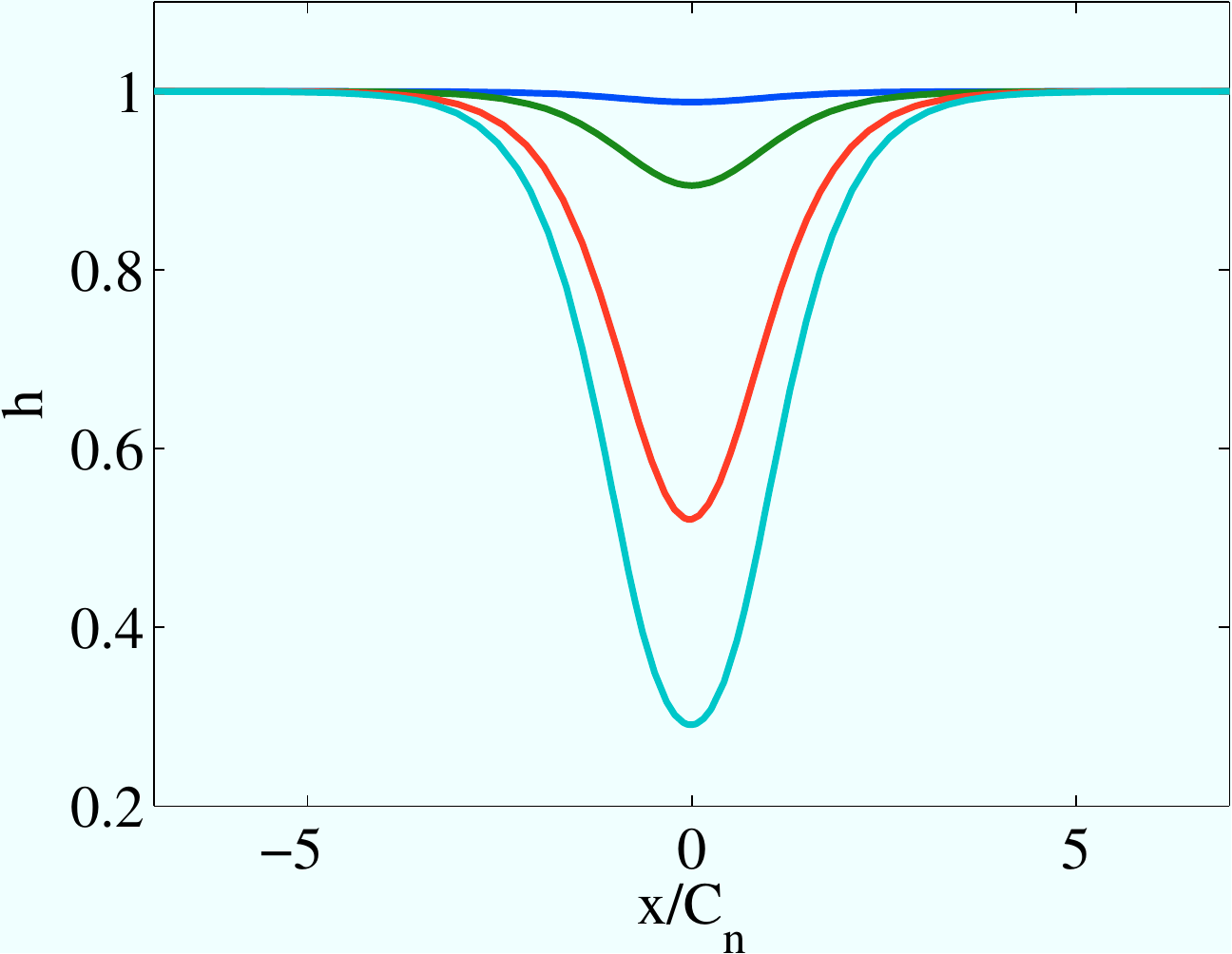}}
\subfigure[]{
\includegraphics[width=0.3\textwidth]{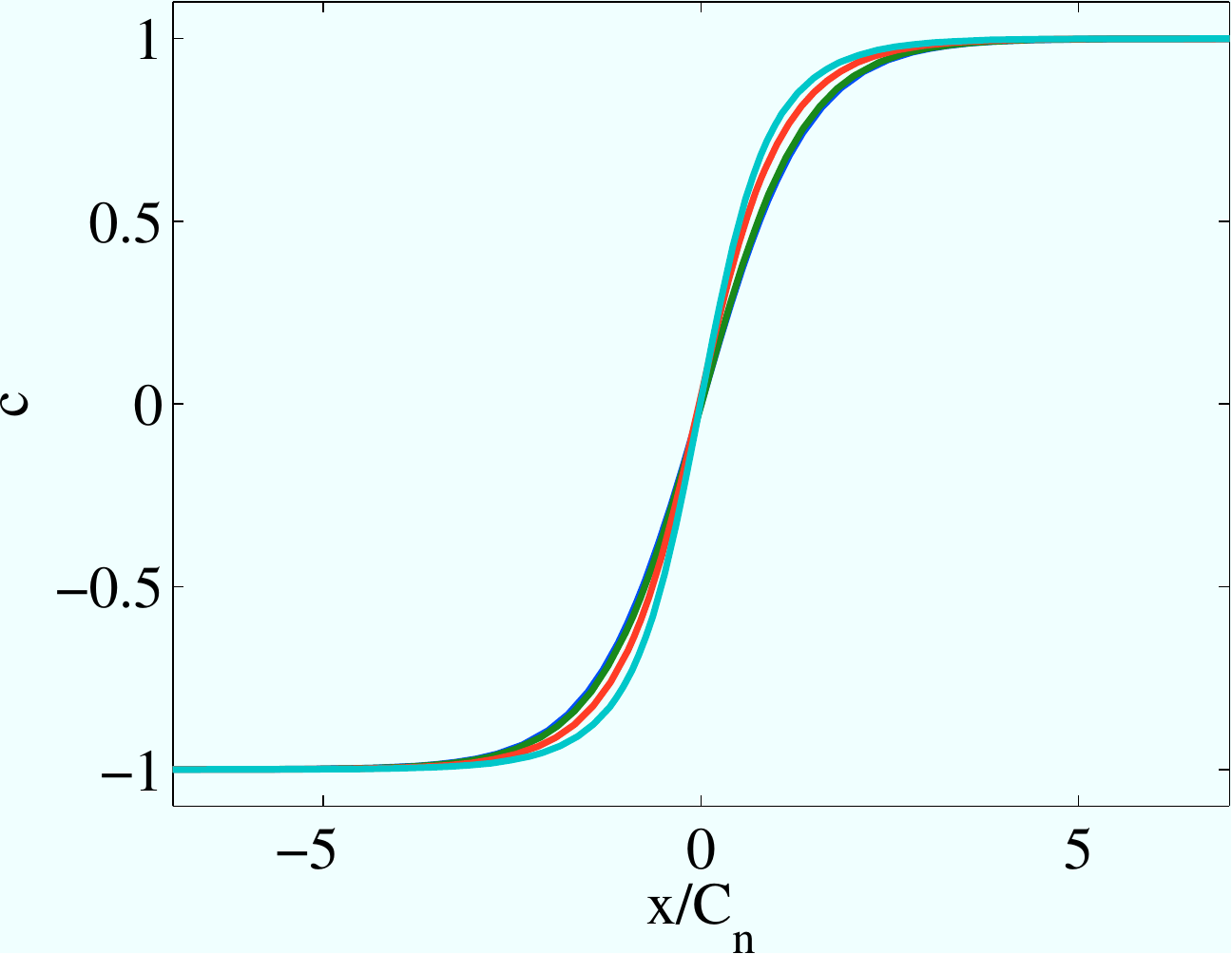}}\\
\subfigure[]{
\includegraphics[width=0.3\textwidth]{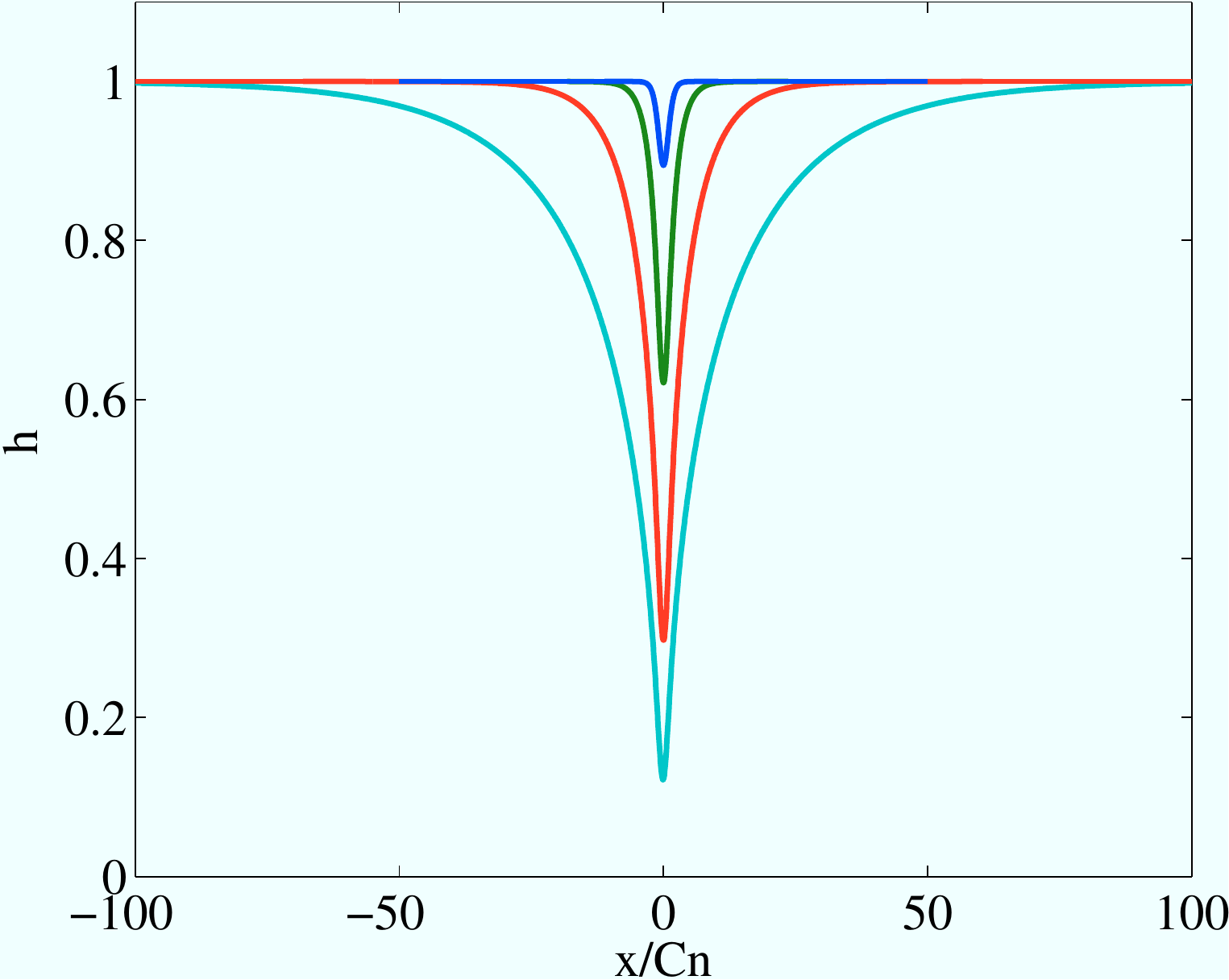}
}
\subfigure[]{
\includegraphics[width=0.3\textwidth]{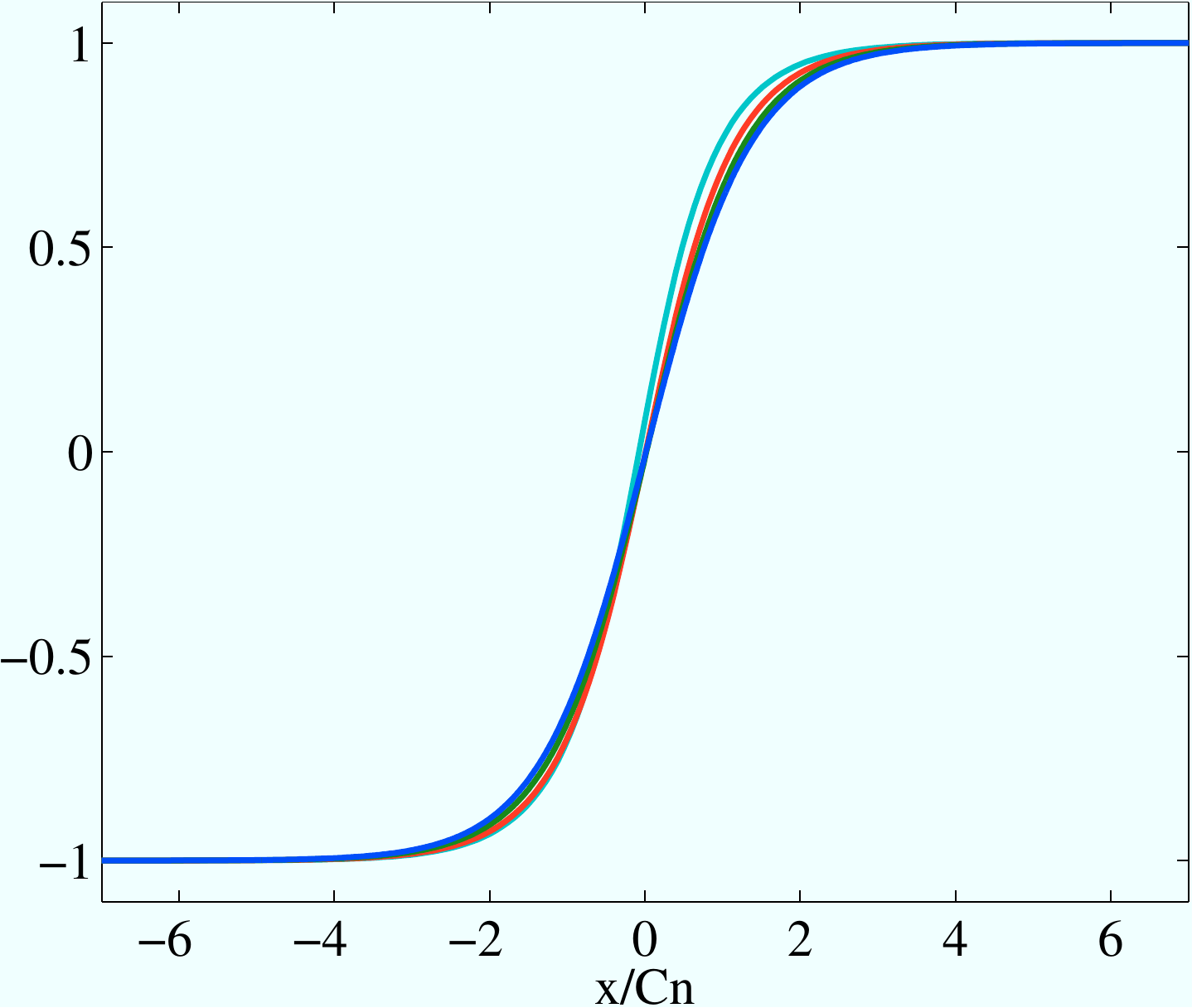}
}
\caption{Solutions of the thin-film equations obtained by solving the boundary-value problem~\eqref{eq:eqm}.
Across the top: the effect on the equilibrium solutions of varying the backreaction strength, for parameter values $C=\Cn^2G=1$
and $r=0.1,1,10,50$.  In (a) the valley deepens with increasing $r$ although
the film never ruptures, while in (b) the front steepens with increasing
$r$.  Figs. (a) and (b) are taken from \'O N\'araigh and Thiffeault~\cite{ONaraigh2007}.
Across the bottom: the effect on the equilibrium solutions of varying the strength of the regularizing potential.  The parameter values are $C=\Cn^2=r=1$
and $G=0.001,0.01,0.1,1$.  In (c) the valley deepens with decreasing $G$ although
the film never ruptures, while in (d) the front steepens with decreasing
$G$.
}
\label{fig:hc}
\end{figure}

\begin{figure}[htb]
\centering
\includegraphics[width=0.4\textwidth]{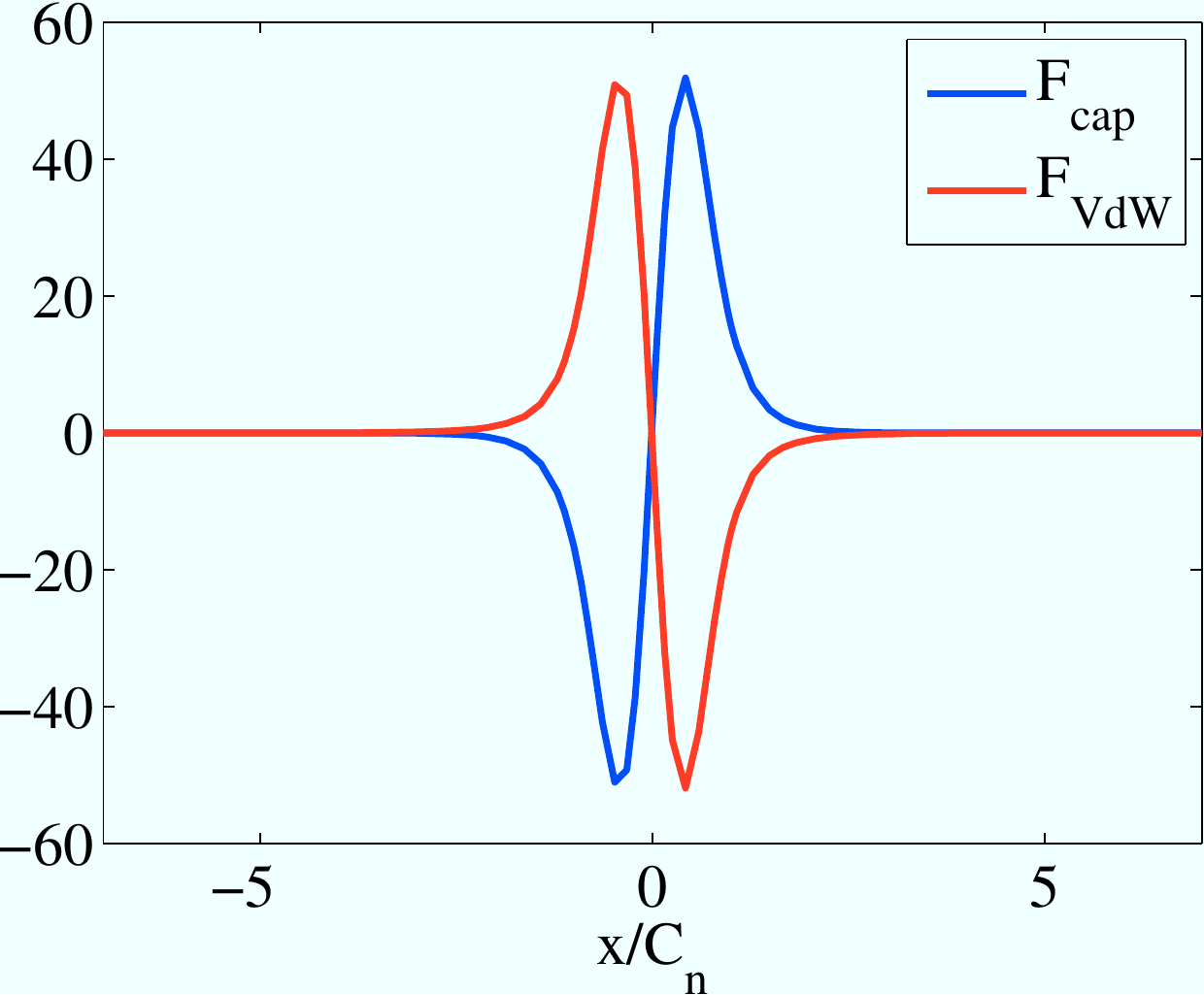}
\caption{Boundary-value solution.  A plot of the forces $F_{\mathrm{cap}}$ and $F_{\mathrm{vdW}}$
for $C=\Cn^2=G=1$ and $r=50$: they have opposite sign; hence, the regularizing potential opposes the rupture-inducing tendency of the backreaction.}
\label{fig:vdw_explain}
\end{figure}

The results for the case with finite surface tension are qualitatively similar.
Here, two parameters characterize the problem: since Eq.~\eqref{eq:eqm1} can be multiplied across by $C$, there are precisely two dimensionless groups, $C \Cn^2G$ and $rC$.  In Fig.~\ref{fig:hc} we present numerical
solutions
exhibiting the dependence of the solutions on these parameters.  As before,
the height field possesses peaks and valleys, where the valleys occur in
the transition region of concentration.  While the valley increases in depth
for
large $r$ or small $G$, rupture never takes place, as guaranteed by the analysis of Sec.~\ref{sec:analysis}.
The repulsive Van der Waals potential therefore has a regularizing effect on
the solutions.  
Indeed, the formation of the valley in the height field has the physical interpretation
of a balance between the Van der Waals and backreaction effects.  From Fig.~\ref{fig:vdw_explain}
we see that the backreaction force, which, through Eq.~\eqref{eq:model} we identify as $F_{\mathrm{cap}}=-rh^{-1}\partial_x\left[h\left(\partial_xc\right)^2\right]$, is of opposite sign to the Van der Waals force $F_{\mathrm{VdW}}=G\partial_x h^{-3}$.  Now in Sec.~\ref{sec:analysis} we showed how the Van der Waals force inhibits rupture; hence, $F_{\mathrm{cap}}$ must promote it.  The depth of the valley in the height field is therefore selected through a balance between rupture-preventing and rupture-promoting effects.

\begin{figure}[htb]
\centering
\subfigure[]{
\includegraphics[width=0.45\textwidth]{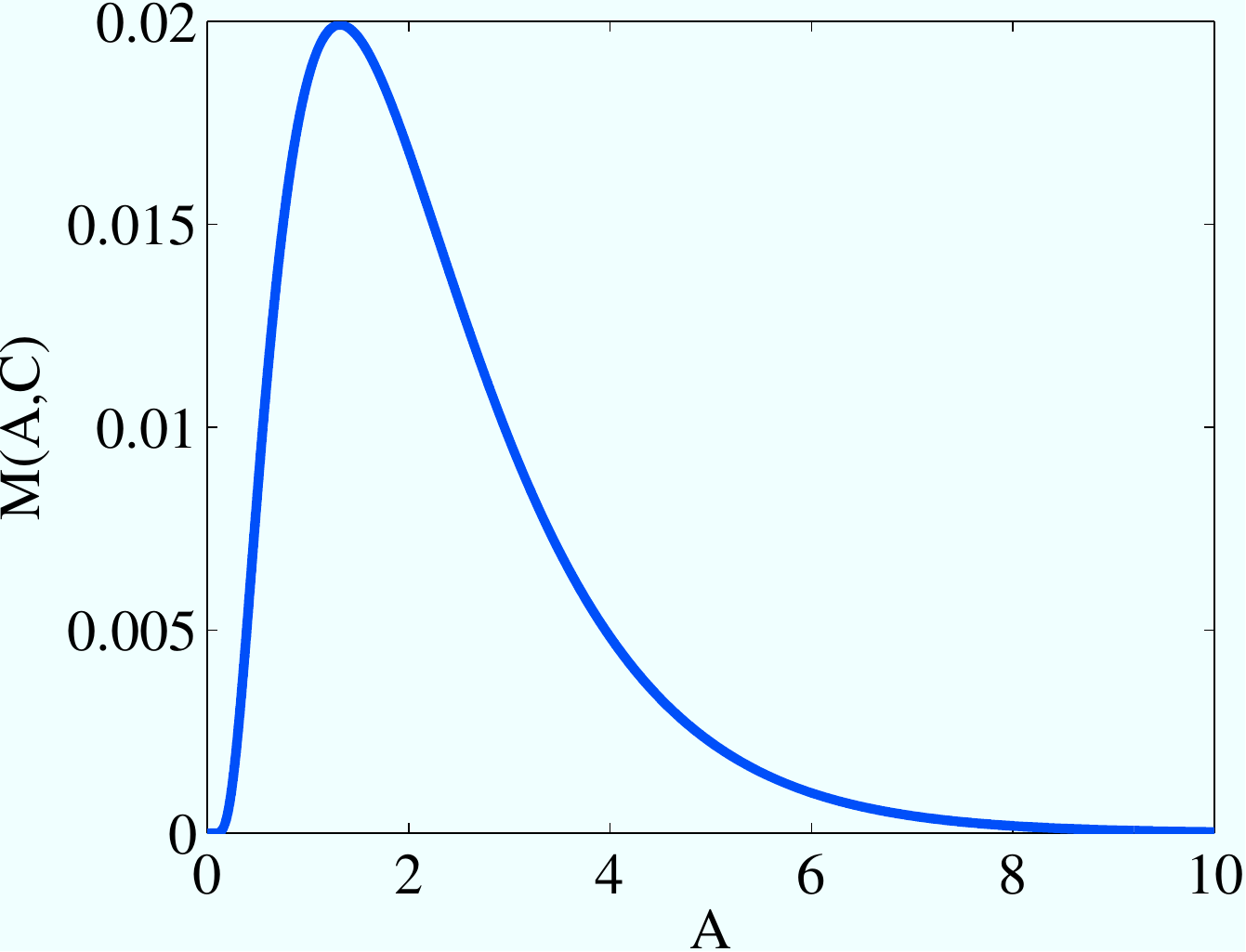}
}
\subfigure[]{
\includegraphics[width=0.45\textwidth]{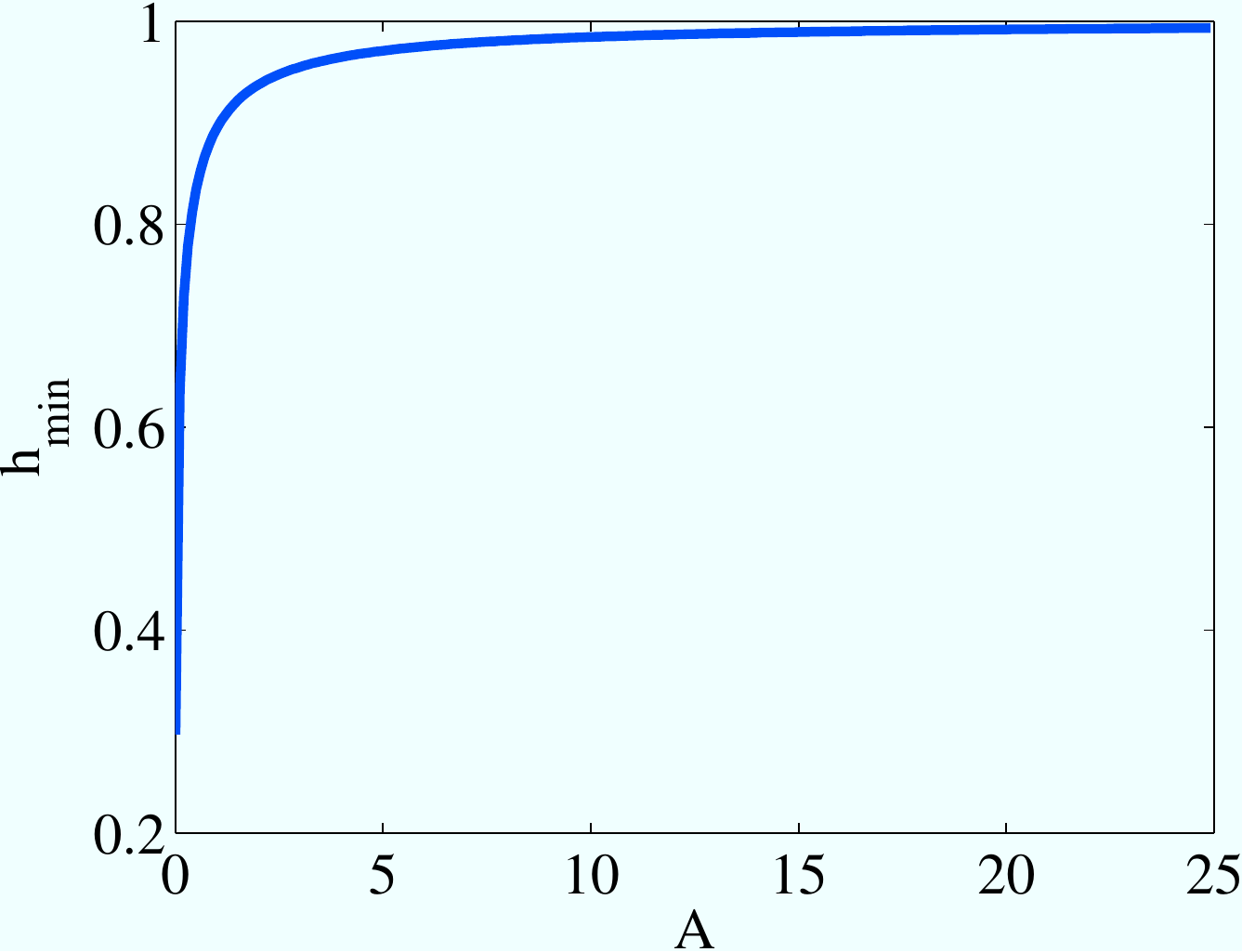}
}
\caption{(a) A typical plot of $s_*\left(G,C\right)$ for $F_0=F_1=\tfrac{1}{2}$
and $C=1$.
 This theoretical lower bound has a different shape from that in (b), which is obtained from a solution of the BVP, with $C=r=1$.  This suggests that while $s_*\left(G,C\right)$ plays an important role
 in the analysis of the model equations, it does not capture the physics
 of film thinning.}
\label{fig:M_A}
\end{figure}

Finally, we compare the results for the minimum free-surface height implied by the solution of the boundary value problem (BVP) with the theoretical lower bound obtained in Sec.~\ref{sec:analysis}.
In terms of the physical parameters of the system, the no-rupture condition
of Sec.~\ref{sec:analysis} is
\[
h_{\mathrm{min}}\geq s_*=\sqrt{2CL(F_0+F_1G)}\left(\sqrt{\frac{e^{4CG^{-1}\left(F_0+F_1G\right)^2}}{e^{4CG^{-1}\left(F_0+F_1G\right)^2}-1}}-1\right)>0,
\]
where $F_1=\tfrac{1}{2}\int_\Omega{dx}\left[h\left(x,0\right)\right]^{-2}\neq0$,
and $F_0=\mathcal{F}\left(0\right)-F_1$.  The function $s_*\left(G,C\right)$ has no
explicit $r$-dependence: although $F_0$ depends on $r$, it is possible to
find initial data to remove this dependence.
We show a representative plot of $s_*\left(G,C\right)$ in Fig.~\ref{fig:M_A}~(a), while in~(b) we show a plot of $h_{\mathrm{min}}$ as a function of $G$, obtained from the solution of the BVP.  Now although these two figures  represent solutions to the model equations for different boundary conditions, a comparison between them is warranted, especially at a domain boundary, where the film thinning is induced by entirely local effects.
The shape of the two bounds in Figs.~\ref{fig:M_A}~(a) and~(b) is different.
Since the bound in Fig.~\ref{fig:M_A}~(b)
is obtained from numerical simulations, and is intuitively correct, we conclude
that it has the correct shape and that the bound of Fig.~\ref{fig:M_A}, while
mathematically indispensable, is not sharp enough to be useful in determining
the parametric dependence of the dip in free-surface height.

\subsection{Numerical studies without a regularizing Van der Waals potential: a study of rupture}
\label{sec:numerics_rupture}

\begin{figure}[htb]
\centering
\subfigure[]{
\includegraphics[width=0.3\textwidth]{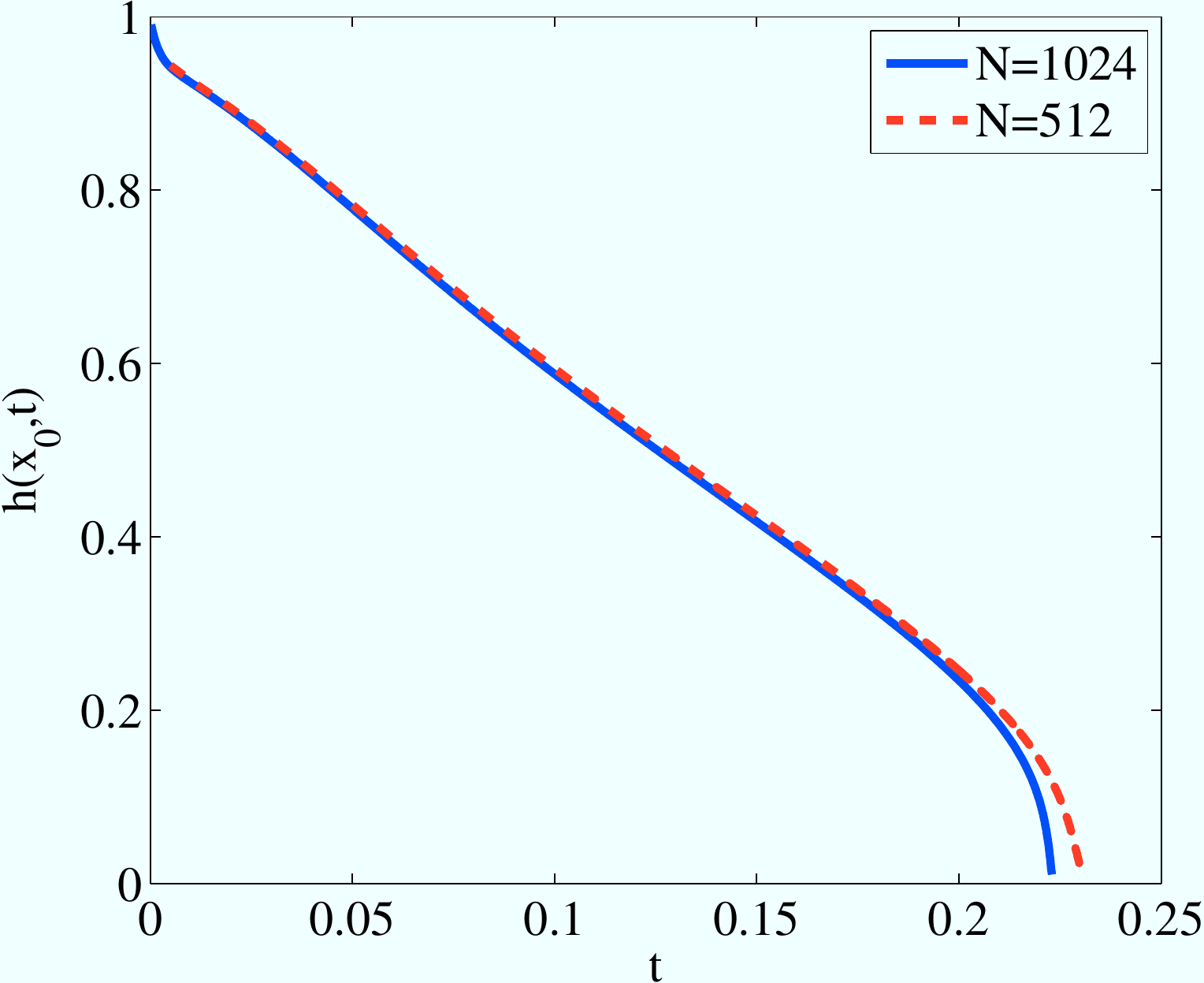}
}
\subfigure[]{
\includegraphics[width=0.3\textwidth]{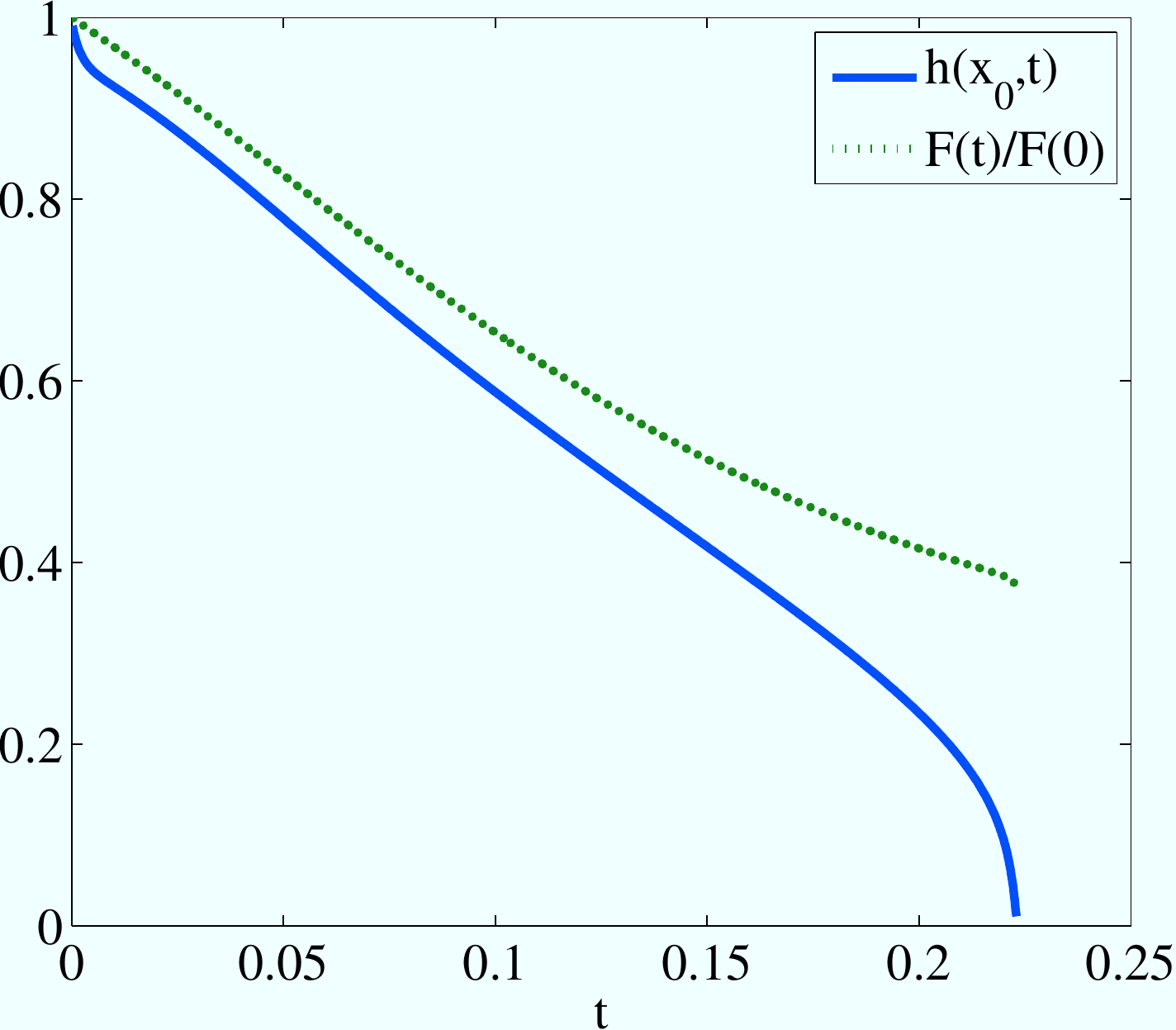}
}
\subfigure[]{
\includegraphics[width=0.3\textwidth]{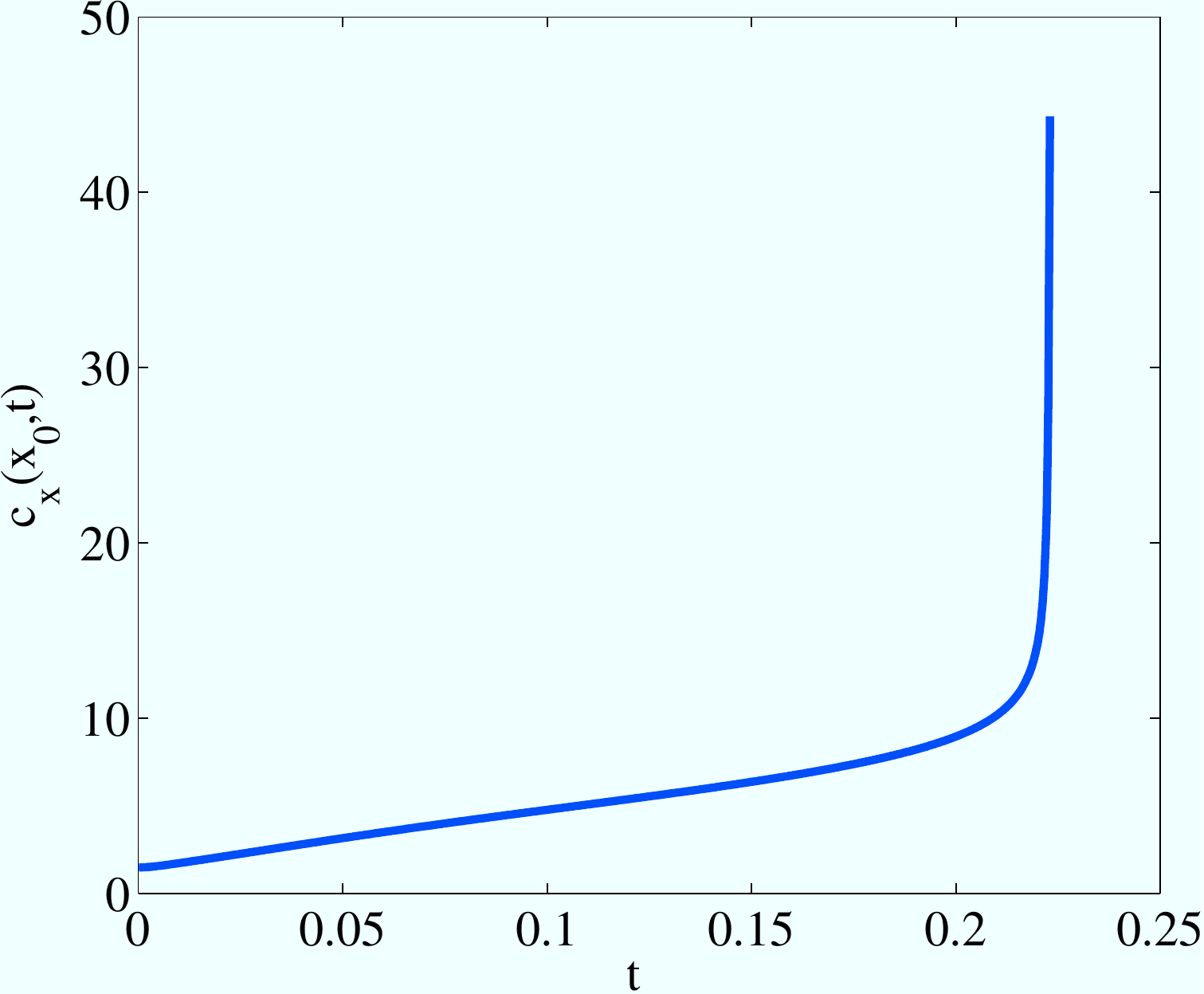}
}
\caption{(a) Temporal evolution of the minimum free-surface height.  The rupture is hastened by grid refinement; (b) our numerical simulation also captures the decay of the free energy; (c)  the rupture coincides with a finite-time singularity, wherein the derivative $c_x$ diverges at the point where $h$ touches down.}
\label{fig:h_min_rupture}
\end{figure}

We perform numerical simulations of the full equations~\eqref{eq:model},
with the following finite-amplitude initial data:
\[
h\left(x,0\right)=1+0.1\sin\left[3\left(2\pi/L\right)x+\pi\right],\qquad c\left(x,0\right)=0.5\sin\left[3\left(2\pi/L\right)x\right].
\]
We use the parameters $r=2$, $G=0$, $C=1$, $\Cn=0.1$, and $L=2\pi$. The calculations are carried out on a periodic domain, with $512$, and $1024$ gridpoints and a timestep $\Delta t=10^{-6}$.   The regularizing Van der Waals force is no longer present, and thus the estimates of Sec.~\ref{sec:analysis} no longer apply.  We therefore examine the possibility that the film will rupture in finite time.  For the parameter values chosen, rupture does indeed occur in finite time, as evidenced by Fig.~\ref{fig:h_min_rupture}.
In the numerical simulation, rupture is only hastened by grid
refinement: this indicates that the effect does not disappear with an
increase in resolution, but is nevertheless difficult to capture precisely.  The simulation also decreases the free energy.  This is consistent with the free-energy decay law derived in Sec.~\ref{sec:analysis}, which relies only on the fact that $h$ should be positive, and holds even for zero Van der Waals forces.  Thus, we are satisfied that the rupture is accurately described by the numerical simulation, and is not simply an artefact.
\begin{figure}[htb]
\centering
\subfigure[]{
\includegraphics[width=0.45\textwidth]{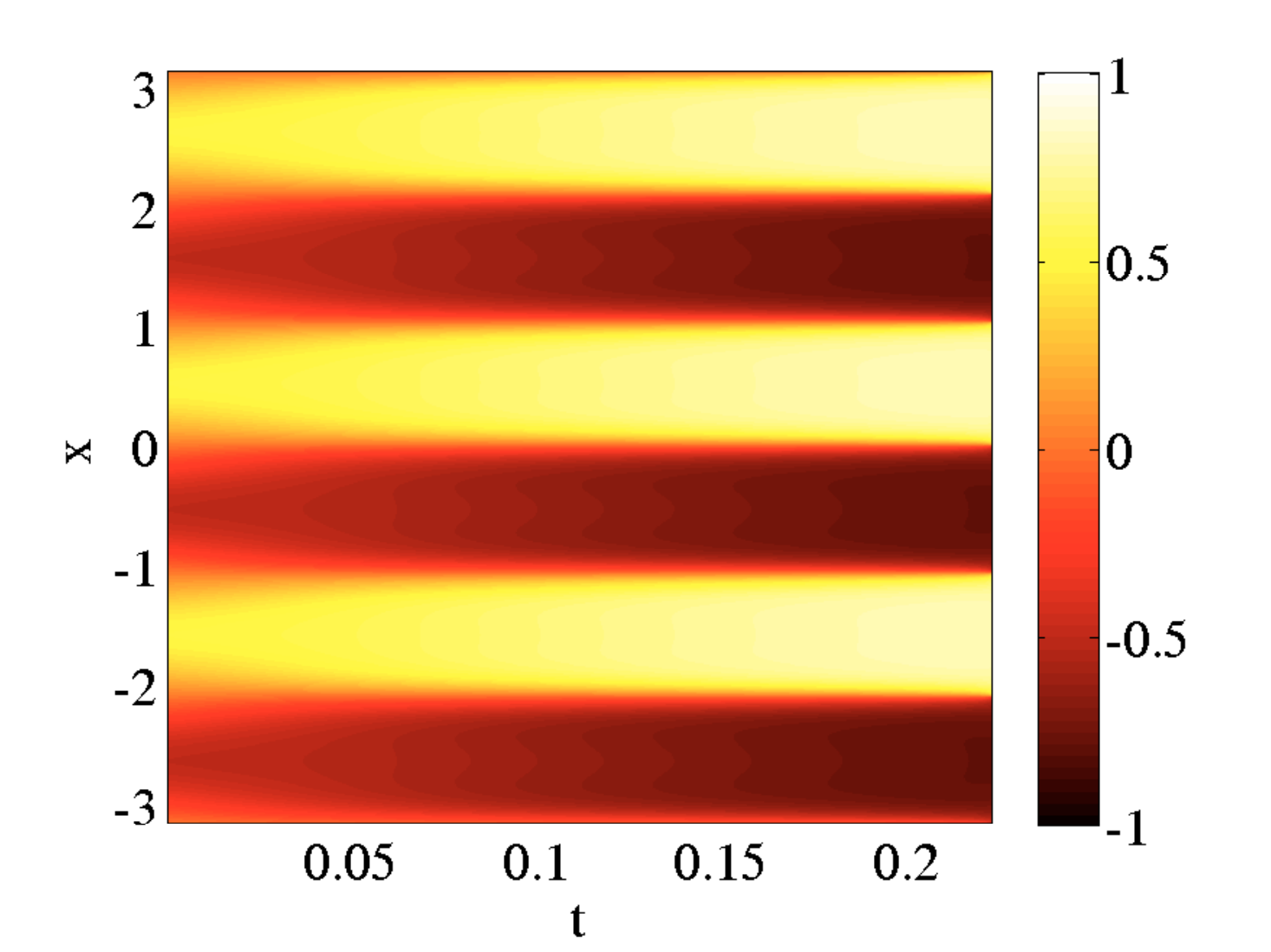}
}
\subfigure[]{
\includegraphics[width=0.45\textwidth]{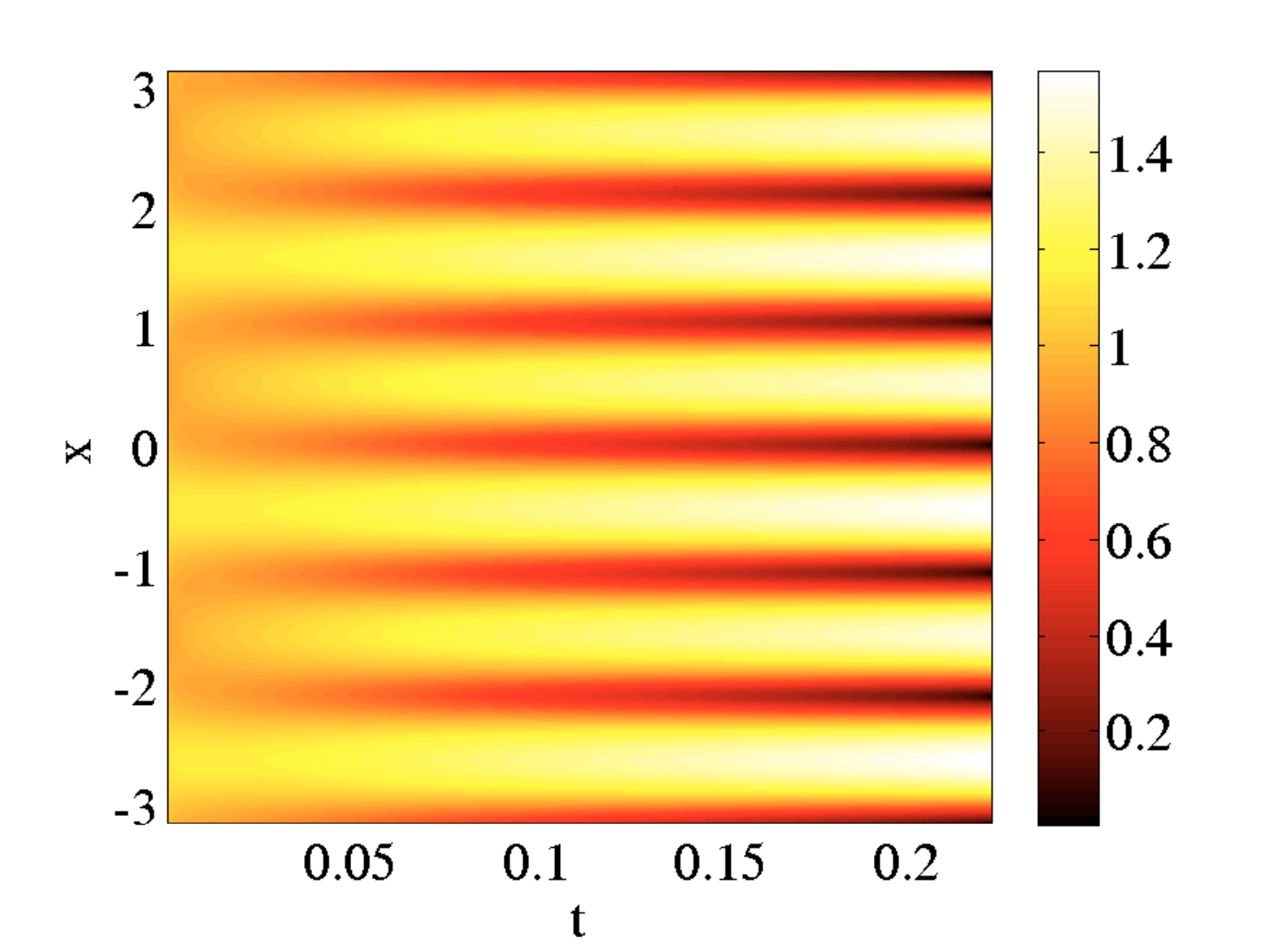}
}
\caption{No regularizing potential and large backreaction effect, $r=2$.  Temporal evolution of (a) the concentration; (b) the free-surface height.  The height touches down to zero in finite time.  A singularity develops in the equations and the gradient of the concentration diverges in the transition zone.}
\label{fig:rupture}
\end{figure}
The evolution towards rupture is shown again in Fig.~\ref{fig:rupture}.  The figure demonstrates yet again that rupture is induced in the transition zones of concentration.
Our understanding of rupture is strengthened by a further examination of the equilibrium case described by Eqs.~\eqref{eq:eqm}. For $G=0$, the $h$-equation reads
\begin{equation}
\frac{1}{C}\frac{\partial^2 h}{\partial x^2}= 
r\left[\tfrac{1}{4}\left(c^2-1\right)^2+\tfrac{1}{2}\left(\frac{\partial{c}}{\partial{x}}\right)^2\right].
\label{eq:eqm0}%
\end{equation}%
As a boundary-value problem, with boundary conditions $h=1$ as
$x\rightarrow\pm\infty$ and $c=\pm1$, $c_x=0$ as $x\rightarrow\pm\infty$, Eq.~\eqref{eq:eqm0} has no solution, since $h_{xx}>0$ everywhere is not compatible with a bounded solution.  In the language of dynamical systems, the proposed boundary conditions for the solution pair $\left(h,h_x\right)$ imply a homoclinic orbit, which is impossible if $h_{xx}$ is positive everywhere.
Thus, the development of a finite-time singularity is consistent with the non-existence of a time-independent solution of the $\left(h,c\right)$ equation pair.

Our numerical study has demonstrated the sharp difference in the two cases wherein the repulsive Van der Waals force is either included or neglected.
In the case where this force is neglected, the film ruptures in finite time, an event that is accompanied by the development of a singularity in the derivative in the concentration.  While interesting mathematically, this is undesirable from a physical point of view.  Consistent with the analysis of Sec.~\ref{sec:analysis}, the inclusion of the Van der Waals term prevents rupture, and enables the development of an equilibrium state.  Note finally that simulations involving two lateral directions have elsewhere been carried
out by the authors~\cite{ONaraigh2007}, and the qualitative features are
similar to those obtained here.

\section{Conclusions}
\label{sec:conclusions}

\noindent Starting from the Navier--Stokes Cahn--Hilliard equations, we have
derived a pair of nonlinear parabolic PDEs that describe the coupled 
effects of phase separation and free-surface variations in a thin film of
binary liquid.  Since we are interested in the long-time outcome of the phase
separation, we have focussed on liquids that experience a repulsive Van der Waals
force, which tends to inhibit film rupture.  Using physical intuition, we
identified a decaying energy functional that facilitated analysis of the
equations.
Based on this decaying energy functional, we have developed a series of \textit{a priori} estimates for positive solutions $\left(h,c\right)$, $h>0$ to the model equations~\eqref{eq:model}.  The H\"older continuity of $h$ obtained through the decaying energy gives rise to a positive lower bound on the height $h$, valid for a general family of repulsive potentials.  These estimates are valid not only in the \textit{a priori} sense described here, but also as a means of demonstrating the \textit{existence} of regular solutions to Eqs.~\eqref{eq:model}, given appropriate initial data~\cite{Friedman1990,ONaraighPhD}.

We carried out one-dimensional numerical simulations of the
equations~\eqref{eq:model} and found that the free-surface height and concentration
tend to an equilibrium state.  The concentration forms domains; that
is, extended regions where $c\approx\pm1$.  The domains are separated
by narrow zones where the concentration smoothly transitions between
the limiting values $\pm1$.  At the transition zones, the free surface
dips below its mean value to form a `valley', a feature of binary thin-film behaviour
that is observed in experiments.  To study the valley depth
as a function of the problem parameters, we focussed on solving the
equilibrium version of Eq.~\eqref{eq:model} as a boundary-value
problem.  This simplification is carried out without much loss of
generality, since our numerical simulations indicate that the system tends asymptotically
to such a state.  We have shown that the valley becomes shallower upon increasing the strength of the repulsive Van der Waals
force, while it deepens  when the backreaction strength is increased.    The film-thinning tendency of the backreaction has been
observed experimentally~\cite{WangH2000, WangW2003, ChungH2004}.
In the limit of zero repulsive Van der Waals forces, the solution of the boundary-value problem implies that the film ruptures, and our temporal numerical simulations confirm this.  Indeed, we have demonstrated that the film ruptures in finite time; simultaneously, the derivative of the concentration becomes singular.  Since such singularities are undesirable from a physical point of view, this result underscores the importance of including Van der Waals forces in studies like this one.

\emph{Acknowledgements.} We thank G.~Pavliotis for helpful discussions.
L.\'O.N. was supported by the Irish government and the UK Engineering
and Physical Sciences Research Council.

\appendix

\section{}

\noindent In this section we outline a numerical technique for the equations
\begin{subequations}
\begin{equation}
\frac{\partial h}{\partial t}+\frac{\partial}{\partial x}\left(hu\right)=0,\qquad
\frac{\partial c}{\partial t}+u\frac{\partial c}{\partial x}=\frac{1}{h}\frac{\partial}{\partial{x}}\left(h\frac{\partial\mu}{\partial{x}}\right),
\end{equation}
where
\begin{equation}
u = - \tfrac{1}{3} h^2 \bigg\{
           \frac{\partial}{\partial{x}}
           \left(-\frac{1}{C}\frac{\partial^2{h}}{\partial{x}^2}
             + \phi\right) + \frac{r}{h} \frac{\partial}{\partial{x}}
           \left[h{\left(\frac{\partial{c}}{\partial{x}}
             \right)}^2\right]\bigg\},
\end{equation}
\begin{equation}
\mu=c^3-c-\frac{\Cn^2}{h}\frac{\partial}{\partial{x}}\left(h\frac{\partial{c}}{\partial{x}}\right).
\end{equation}%
\label{eq:appendix_model_intro}%
\end{subequations}%
We impose periodic boundary conditions on the solution $\left(h,c\right)$ and its derivatives.  The solution $\left(h,c\right)$ is discretized on a
regular spatial grid such that the vector pair $\left(\bm{h}\left(t\right),\bm{c}\left(t\right)\right)\in \mathbb{R}^N\times\mathbb{R}^N$ represents the discretized solution at time $t$.  The derivatives are approximated as centred finite differences with the periodic boundary conditions taken into account.  Thus, the derivatives are reduced to matrix operators $\mathcal{D}_j$ on $\mathbb{R}^N$ (here the subscript `$j$' denotes the order of the derivative).  The solution is marched forwards in time using a semi-implicit Euler algorithm,
\begin{eqnarray}
\frac{\bm{h}^{n+1}-\bm{h}^{n}}{\Delta t}&=&-\frac{1}{3C}\mathcal{D}_1\left[\left(\bm{h}^n\right)^{\cdot3}\cdot\left(\mathcal{D}_3\bm{h}^{n+1}\right)\right]+\bm{S}_h^n,\nonumber\\
\frac{\bm{c}^{n+1}-\bm{c}^{n}}{\Delta t}&=&-\Cn^2\left[\mathcal{D}_4\bm{c}^{n+1}\right]+\bm{S}_c^n,
\label{eq:ode_solve}
\end{eqnarray}
where the `dot' is used here to denote pointwise multiplication, $\bm{x}\cdot\bm{y}=\left(x_1y_1,\cdots,x_Ny_N\right)$, $\bm{x}^{\cdot a}=\left(x_1^a,\cdots,x_N^a\right)$, and where the `source' terms $\bm{S}_h$ and $\bm{S}_c$ are defined as follows: %
\begin{eqnarray}
\bm{S}_h&=&  \tfrac{1}{3}\mathcal{D}_1 \bigg\{\left(\bm{h}^{\cdot3}\right)\cdot \bigg[
          \mathcal{D}_1\bm{\phi}
                 + r \left(\bm{h}^{\cdot-1}\right)\cdot\left[\mathcal{D}_1
           \left(\bm{h}\cdot\left(\mathcal{D}_1\bm{c}\right)^{\cdot2}\right)\right]\bigg]\bigg\},\nonumber\\
\bm{S}_c&=& -\bm{u}\cdot\left(\mathcal{D}_1\bm{c}\right)+\left(\bm{h}^{\cdot-1}\right)\cdot\left(\mathcal{D}_1\bm{h}\right)\cdot\left(\mathcal{D}_1\bm{\mu}\right)+\mathcal{D}_2\left[\bm{c}^{\cdot3}-\bm{c}-\Cn^2\left(\bm{h}^{\cdot-1}\right)\cdot\left(\mathcal{D}_1\bm{h}\right)\cdot\left(\mathcal{D}_1\bm{c}\right)\right],\nonumber\\
\bm{u}&=&-\tfrac{1}{3} \left(\bm{h}^{\cdot2}\right)\cdot \bigg[
          -\frac{1}{C} \mathcal{D}_3\bm{h}+
          \mathcal{D}_1\bm{\phi}
                 + r \left(\bm{h}^{\cdot-1}\right)\cdot\left[\mathcal{D}_1
           \left(\bm{h}\cdot\left(\mathcal{D}_1\bm{c}\right)^{\cdot2}\right)\right]\bigg],\nonumber\\
\bm{\mu}&=&\bm{c}^{\cdot3}-\bm{c}-\Cn^2\left(\bm{h}^{\cdot-1}\right)\cdot\left(\mathcal{D}_1\bm{h}\right)\cdot\left(\mathcal{D}_1\bm{c}\right)-\Cn^2\mathcal{D}_2\bm{c}.
\end{eqnarray}

Equations~\eqref{eq:ode_solve} are re-written as
\begin{eqnarray}
\left[\bm{1}+\frac{\Delta t}{3C}\mathcal{D}_1\left(\left(\bm{h}^n\right)^{\cdot3}\cdot\mathcal{D}_3\right)\right]\bm{h}^{n+1}&=&\bm{h}^{n}+\Delta t\,\bm{S}_h^n,
\label{eq:ode_solve1}\\
\left[\bm{1}+\Delta t \Cn^2\mathcal{D}_4\right]\bm{c}^{n+1}
&=&\bm{c}^{n}+\Delta t\,\bm{S}_c^n,
\label{eq:ode_solve2}
\end{eqnarray}
a linear problem that is solvable for $\left(\bm{h}^{n+1},\bm{c}^{n+1}\right)$ by matrix inversion.  The $\bm{h}$-equation manifestly conserves the sum $\sum_{i=1}^N h_i$, since $\sum_{i=1}^N\sum_{j=1}^N\mathcal{D}_{1,ij}v_j=0$, for any vector $\bm{v}\in\mathbb{R}^N$.  Other semi-implicit numerical schemes
\begin{figure}[htb]
\centering
\includegraphics[width=0.5\textwidth]{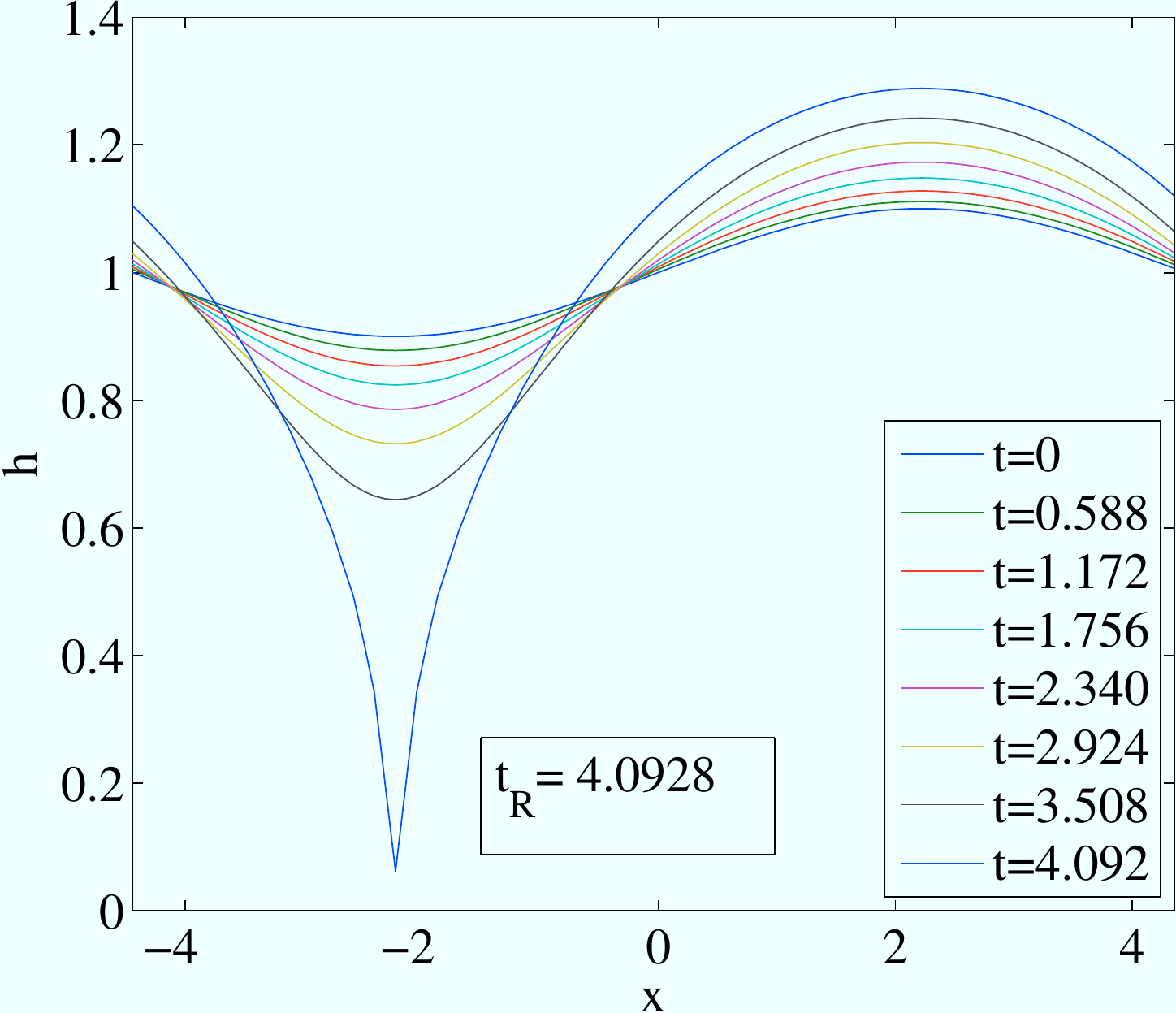}
\caption{Comparison with the work of Burelbach \textit{et al.} for the rupture of a single-component fluid under the influence of an attractive Van der Waals potential.  The rupture happens in finite time and is calculated here as $t_R=4.093$, for simulation parameters $\Delta t=10^{-5}$ and $N=100$.}
\label{fig:burelbach}
\end{figure}
that only approximate this conservation law can fail near rupture ($h\rightarrow0$). %
 One such technique is the otherwise successful method of Kondic~\cite{Kondic2003}, which replaces the term $\left(\bm{h}^n\right)^{\cdot3}\cdot\left(\mathcal{D}_3\bm{h}^{n+1}\right)$ with $\left(\bm{h}^{n+1}\right)^{\cdot3}\cdot\left(\mathcal{D}_3\bm{h}^{n+1}\right)$ and th\emph{}us involves a Newton iteration at each timestep.

Now although the implied conservation law $\partial_t\int_0^L ch\,dx$ is not manifest in the $\bm{c}$-equation~\eqref{eq:ode_solve2}, we have verified that a sufficiently small stepsize and gridsize guarantees its conservation in practice.  The implicit step in Eq.~\eqref{eq:ode_solve2} is particularly fast because the matrix $1+\Delta t \Cn^2 \mathcal{D}_4$ need only be inverted once.
This implicit treatment of the high-order derivatives in Eqs.~\eqref{eq:ode_solve1}--\eqref{eq:ode_solve2} also ensures numerical stability for large timesteps that would otherwise cause numerical blowup.

We verify the correctness of our numerical scheme by comparing it against a set of well-known results for the single-component equation
\[
h_t+\left(h^{-1}h_x\right)_x+\left(h^3 h_{xxx}\right)_x=0,
\]
which touches down to zero in finite time.  This equation has been studied by Burelbach \textit{et al.}~\cite{Burelbach1988}, with the initial condition
\[
h_0=1+0.1\sin\left(x/\sqrt{2}\right).
\]
They observed finite-time rupture and estimated the 
the rupture time as $t_R=4.164$, based on a numerical study with $\Delta t=10^{-5}$ and $N=40$.  With these grid parameters, the numerical scheme~\eqref{eq:ode_solve1}--\eqref{eq:ode_solve2} gives a rupture time $t_R=4.145$.  A possible source for the small discrepancy of estimates is the fact that we have kept $\Delta t = 10^{-5}$ for the duration of the simulation; Burelbach \textit{et al.} refine it as rupture approaches, until $\Delta t=10^{-5}$.
  Refining the grid ($N=100$) gives a reduced rupture time $t_R=4.093$ (see Fig.~\ref{fig:burelbach}), a reduction that is consistent with Fig.~3 of Burelbach \textit{et al.}  In conclusion, this test of our scheme validates its applicability to the two-component equations~\eqref{eq:appendix_model_intro}.

%\bibliographystyle{unsrt}
%\bibliography{thin_film_bibliography}

\end{document}